\documentstyle[epsfig,11pt]{article}
\newcommand{\be}{\begin{equation}}
\newcommand{\ee}{\end{equation}}
\newcommand{\ben}{\begin{eqnarray}}
\newcommand{\een}{\end{eqnarray}}
\newcommand{\bc}{\begin{center}}
\newcommand{\ec}{\end{center}}

\begin{document}

\title{Statistical mechanics and the description of the early
universe\\ I.: Foundations for a slightly non-extensive cosmology}

\author{M. E. Pessah$^{a,b}$, Diego F. Torres$^{b}$ and H.
Vucetich$^{a}$
\\
{\small $^a$Facultad de Ciencias Astron\'omicas y Geof\'{\i}sicas,
UNLP, Paseo del Bosque s/n, 1900, La Plata, Argentina}
 \\ {\small $^b$Instituto Argentino de
Radioastronom\'{\i}a, C.C.5, 1894 Villa Elisa, Buenos Aires,
Argentina} }

\date{\today }

\maketitle

\begin{abstract}

We analyze how the thermal history of the universe is influenced
by the statistical description, assuming a deviation from the
usual Bose-Einstein, Fermi-Dirac and Boltzmann-Gibbs distribution
functions. These deviations represent the possible appearance of
non-extensive effects related with the existence of long range
forces, memory effects, or evolution in fractal or multi-fractal
space. In the early universe, it is usually assumed that the
distribution functions are the standard ones. Then, considering
the evolution in a larger theoretical framework will allow to test
this assumption and to place limits to the range of its validity.
The corrections obtained will change with temperature, and
consequently, the bounds on the possible amount of non-extensivity
will also change with time.  We generalize results which can be
used in other contexts as well, as the Boltzmann equation and the
Saha law, and provide an estimate on how known cosmological bounds
on the masses of neutrinos are modified by a change in the
statistics. We particularly analyze here the recombination epoch,
making explicit use of the chemical potentials involved in order
to attain the necessary corrections. All these results constitute
the basic tools needed for placing bounds on the amount of
non-extensivity that could be present at different eras and will
be later used to study primordial nucleosynthesis.

\end{abstract}

\bigskip


\newpage



\section{Introduction}

This paper is the first in a series that will thoroughly analyze
the influence of the statistical description in the standard
picture we have for the evolution of the early universe. To do so,
we shall insert the usual statistical mechanics, i.e.
Boltzmann-Gibbs', in a larger framework, given by non-extensive
theories. These theories are parameterized by a new degree of
freedom, related to the amount of non-extensivity present in the
system under consideration. The standard statistical mechanics is
the only one that respects the property of extensivity, i.e. the
entropy of a system formed by several subsystems equals the sum of
the entropies of each separate subsystem. Then, considering this
case as a particular situation out of a larger set of
possibilities, our aim will be twofold. On the one hand, we shall
comment on the reasons by which the universe as a whole could
deviate from being an extensive system, and on the possible amount
of these deviations. On the other, we shall systematically analyze
how many of the standard results of the usual cosmological model,
the hot big bang theory, get affected by such an slight change.
This change will manifest itself in the form of the quantum
distribution functions, that we shall modify at the beginning and
use thorough the paper. Using these generalized predictions, we
shall be in position to assess how our present knowledge of
observational data can bound the statistics operative at different
eras of cosmic evolution. The corrections obtained will change
with temperature, and consequently, the bounds on the possible
amount of non-extensivity will also change with time.

We shall of course draw upon some previous results, for instance,
the modification that a change in the statistics introduce in the
energy density of the universe was previously studied by several
authors. Particularly, some of us have been working in the topic
of nucleosynthesis, under what it is known as the asymptotic
approach of quantum distribution functions \cite{PRL,PHYSICA}.
However, we shall re-derive some of these results in order to have
all of them within the same non-extensive framework; this will
allow to consistently use our new results in the follow up
applications. Indeed, these preliminary works will be used as the
launch pad for a consistent analysis on the influence of
statistical mechanics on all standard -textbook- cosmology: We
shall generalize and obtain corrections due to this new setting
(up to first order in the deviation parameter) for processes going
from the decoupling of hot and cold relic species,
nucleosynthesis, recombination, to matter-radiation equality. In
our way up, we shall as well derive results which can be used in
other contexts, as the generalized Boltzmann equation and the Saha
law. A direct application of them is also carried on here,
particularly in the section concerning neutral hydrogen formation.

Many are the papers on astrophysics and cosmology using
non-standard statistical mechanics. Among them, we would like to
mention here the works related with the solar core \cite{SOLAR},
where an interesting analysis of neutrino production in a
non-extensive setting was presented. It is both interesting and
instructive to follow the recent discussion on the need for a new
statistical description in the solar core, to that end see the
papers of Refs. \cite{BAC,RESP}. Another astrophysical application
has been on the topic of high and ultra-high energy cosmic rays
\cite{CR}. In particular, it was recently proven that a
non-extensive setting can not solve the GZK cutoff problem
\cite{Q-RAYS}. Of course we must recall other previous papers on
cosmology \cite{COSMOLOGY}, and make a special remark on the proof
of the $T \propto R^{-1}$ relationship \cite{BARRACO}, the
analysis of COBE satellite data as a bound for non-extensivity
\cite{Tsallis3}, the use of precision cosmology (MAP, Planck,
SDSS) with the same aim \cite{TORRES-PHYSA}, and some previous
ideas on kinetic theory \cite{BARRACO-2}. This series of papers
encompass most of these results, enlarge them, and tries to form
an unified picture of non-extensive cosmology, both bounding its
possible range of validity and extent and, from a formal point of
view, studying how a generally non-stated assumption can hide lots
of implications.

The layout of this paper is as follows. In the sake of making a
self-contained work, especially for cosmologists, we shall state
in Section 2 some of the most important ideas and the basic
development of non-extensive statistics. For the ease of the
discussion, some brief historical comments will be made also
there. We shall also present the new quantum distribution
functions, that we shall use in the rest of the work, following in
this regard the discussion made by B\"{u}y\"{u}kk{\i}l{\i}\c{c} et
al. and T{\i}rnakl{\i} et al. \cite{TUR}. Section 2, then, does
not pretend to show new results. Section 3 discusses the values of
the plasma parameter along cosmic history, and assess the a priori
possibility that the universe as a whole can be considered as a
non-extensive system. Section 4 and 5 are the basis for all
further development; they state the thermodynamical picture and
the cosmologically conserved quantities in this new setting. Then,
some new applications follow: Section 6 is concerned with the
decoupling of relativistic and non-relativistic species; Section
7, with the process of recombination; and Section 8, with the new
Boltzmann equation and the study of the process of freezing. Then,
a brief Section 9 states the predictions for the current values of
some observables of cosmological importance; and Section 10
studies hot and cold relics in the new framework and generalizes
some of the well known bounds on neutrino masses and on other
species. Finally, in Section 11, it is briefly accounted for the
correction that non-extensivity would introduce to the value of
matter-radiation time and temperature. We end by summarizing and
obtaining some conclusions of a general nature. Paper II in this
series will be an analytical analysis of the nucleosynthesis
process, going all the way up to the formation of Deuterium and
Helium 4. We shall be specially concerned with the analysis of the
nuclear principle of detailed balance and the influence it has in
this new setting.

\section{Basics of non-extensive statistics}

Among physicist and astronomers, there is a non-stated consensus
on that the Boltzmann-Gibbs (BG) statistical mechanics is always
applicable. However, in analogy with Newtonian dynamics and
Special Relativity, it could a priori justified to consider it
just as part of a bigger framework, an enlarged statistical
description where extensive as well as non-extensive phenomena
could be taken into account \cite{BRAZ}. An interesting
generalization of the BG entropy form has been recently proposed
by Tsallis \cite{t2} (for recent reviews see
\cite{Tsallis1,Tsallis2}, for a full bibliography see \cite{WEB}).
This new entropy, that we discuss below, possesses the usual
properties of positivity, concavity and irreversibility, and
generalizes the additivity in a non-extensive way. Examples in
which BG statistics seems to present serious problems are systems
for which there are long-range forces, and/or which present memory
effects, and/or are subject to evolution in a non-Euclidean space
\cite{sistemasconproblemas}.

The generalized form for the entropy is
\begin{equation}
\label{eq:Sq} S_q= k \frac{1-\sum_{i=1}^{W} p_i^q}{q-1},\qquad  q
\in {\bf R},
\end{equation}
where $k$ a positive constant, $W$ the total number of physical
states accessible to the system, and the set of probabilities
$p_i$ satisfies
\begin{equation}
\sum_{i=1}^{W} p_i =1.
\end{equation}
Eq. (\ref{eq:Sq}) recovers the usual (BG) form for the entropy in
the limit $q\rightarrow 1$, i.e.
\begin{equation}
\lim_{q\rightarrow 1} S_q=-k \sum_{i=1}^{W} p_i \ln{p_i}.
\end{equation}
The entropic index $q$, intimately related to, and determined by,
microscopic dynamics, characterizes the system under
consideration. This reflects itself in the pseudo-additivity law
for the entropy $S_q$,
\begin{equation}
  \label{eq:SqAB} S_q(A+B)/k = S_q(A)/k + S_q(B)/k + (1-q)
[S_q(A)/k][S_q(B)/k],
\end{equation}
where $A$ y $B$ are two independent systems in the sense that the
joint probabilities (those corresponding to the system $A+B$) are
such that $p_{ij}(A+B)=p_i(A)p_j(B)$.

\subsection{Expectation values}

We now introduce the following non-normalized expectation value,
\begin{equation}
\langle A\rangle_q\equiv\sum_{i=1}^{W} p^q_i A_i
\end{equation}
such that $\langle A\rangle_1$ corresponds to the standard mean
value for the observable $A$. If the system is a quantum one, its
description is given in terms of the density operator, $\rho$,
with eigenvalues $\{p_i\}$. Then, the generalized entropy is
\begin{equation}
\label{eq:Sq3} S_q=k\frac{1-Tr \rho^q}{q-1} \qquad (Tr\rho=1),
\end{equation}
and the expectation value becomes
\begin{equation}
\langle A\rangle_q\equiv Tr \rho^q A.
\end{equation}
Eq. (\ref{eq:Sq3}) can be recast as
\begin{equation}
\label{eq:Sq4} S_q=-k\langle \ln_q \rho \rangle_q.
\end{equation}
If the system is classic, and the relevant variables are continuum
ones, we can describe it by a probability distribution
$p(\vec{r})$, where $\vec{r}$ is a dimensionless variable, say, in
a many-body phase space. Then, the generalized entropy will be
\begin{equation}
\label{eq:Sq6} S_q=k\frac{1-\int d\vec{r}[p(\vec{r})]^q}{q-1}
\qquad {\rm with} \int d\vec{r}p(\vec{r})=1 ,
\end{equation}
and the expectation value,
\begin{equation}
\langle A \rangle_q\equiv \int d\vec{r}[p(\vec{r})]^q A(\vec{r}).
\end{equation}
We shall present the formalism in the case in which the system is
described by a set of microscopic probabilities, $W$.

\subsection{Canonical Ensemble}

The first non-trivial physical situation is that of a system in
thermal contact with a thermostat at a temperature $T$. We must
then extremize the entropy taking into account several
constraints. The first one is just the definition of probability,
\begin{equation} \label{eq:piconstrain1} \sum_{i=1}^W p_i=1.
\end{equation} For the rest, a detailed discussion is in order.

\subsubsection{Internal energy}

There is some freedom concerning the choice of the constraint
imposing the relationship among the different energy levels with
the total internal energy. The first choice was introduced by
Tsallis \cite{t2}: to Eq. (\ref{eq:piconstrain1}), it is added the
following constraint
\begin{equation}
\label{eq:Econstrain1} \sum_{i=1}^W p_i \epsilon_i =U^{(1)}.
\end{equation}
Here, an index $(1)$ refers to the first choice, and
$\{\epsilon_i\}$ are the Hamiltonian eigenvalues of the system.
Using standard techniques, it can be seen that the
$\{p_i\}$-values that extremize $S_q$ with the imposed constraints
are
\begin{equation}
p_i^{(1)}=\frac{[1-(q-1)\beta^{\star}\epsilon_{i}]^{1/(q-1)}}
{\sum_{j=1}^W[1-(q-1)\beta^{\star}\epsilon_{j}]^{1/(q-1)}}.
\end{equation}
It must be said that $\beta^{\star}$ is not the Lagrange
multiplier associated with the constraint upon the internal
energy. This expression recovers the usual one $(p_i
  \propto e^{-\beta \epsilon_i})$ in the limit $q\rightarrow 1$
  and it depends on the microscopic energies as a power law
  instead of the familiar exponential function.

The second choice \cite{t2} postulates that
\begin{equation}
\label{eq:Econstrain2} \sum_{i=1}^W p_i^q \epsilon_i =U^{(2)}.
\end{equation}
The $\{p_i\}$-values that now extremize $S_q$ are
\begin{equation}
\label{eq:pi2}
p_i^{(2)}=\frac{[1-(1-q)\beta\epsilon_{i}]^{1/(1-q)}} {Z_q^{(2)}},
\end{equation}
where we have defined the generalized partition function
\begin{equation}
\label{eq:Z2}
Z_q^{(2)}=\sum_{j=1}^W[1-(1-q)\beta\epsilon_{j}]^{1/(1-q)}.
\end{equation}
This result differs from the previous in that the role played by
$(1-q)$ is now equivalent to what was played by $(q-1)$, and that
in this case, $\beta$ is the Lagrange multiplier associated with
the constraint on the internal energy. The probability
distribution can be conveniently recast as
\begin{equation}
p_i^{(2)}=\frac{e_q^{-\beta \epsilon_i}}{Z_q^{(2)}} \qquad {\rm
with} \qquad  Z_q^{(2)}\equiv \sum_{j=1}^W e_q^{-\beta \epsilon_i}
,
\end{equation}
what allows to prove the following series of equalities
\cite{Curado},
\begin{eqnarray}
\frac{1}{T} &=& \frac{\partial S_q}{\partial U_q^{(2)}},\\
F_q^{(2)} &\equiv& U_q^{(2)}-T S_q=-\frac{1}{\beta} \ln_q
Z_q^{(2)}, \\ U_q^{(2)} &=& -\frac{\partial \ln_q
Z_q^{(2)}}{\partial \beta},\\ C_q^{(2)} &\equiv& T \frac{\partial
S_q}{\partial T} = \frac{\partial U_q^{(2)}}{\partial T} = -T
\frac{\partial^{2}F_q^{(2)}}{\partial T^2},
\end{eqnarray}
with $F_q, U_q$ and $C_q$ standing for the corresponding
generalizations of Helmholtz free energy, internal energy and
specific heat. Then, the formal thermodynamical structure
(Legendre transformations) remains valid. Three unwanted
consequences, however, should be noted \cite{Tsallis5}
\begin{enumerate}
\item The distribution given by Eqs. (\ref{eq:pi2}) and
(\ref{eq:Z2}) is not invariant against an uniform shift of the
energy zero.
\item The mean value of a constant differs from the constant itself,
$\langle 1\rangle_q \neq $1.
\item Finally, if two systems $A$ and $B$ are such that
$p_{ij}^{A+B}=p_i^A p_j^B$ and $\epsilon_{ij}^{A+B}=\epsilon_i^A +
\epsilon_j^B$, then
\begin{equation}
\label{eq:U2qAB} U_q^{(2)}(A+B)/k = U_q^{(2)}(A)/k +
U_q^{(2)}(B)/k + (1-q)
[U_q^{(2)}(A)S_q(B)/k][U_q^{(2)}(B)S_q(A)/k],\end{equation} what
differs from $U_q^{(2)}(A)/k + U_q^{(2)}(B)/k $. This means that
the energy is not an additive quantity.
\end{enumerate}

A third choice for the energy constraint was proposed by Plastino,
Mendes and Tsallis \cite{Tsallis5}, and its key aspect is
normalization. The constraint is now,
\begin{equation}
\label{eq:Econstrain3} \frac{\sum_{i=1}^W p_i^q
\epsilon_i}{\sum_{i=1}^W p_i^q} =U^{(3)},\end{equation} i.e. it
weights the eigenvalues with a set of escort probabilities
$p_i^q/\sum_{i=1}^W p_i^q$. The set $\{ p_i \}$ which now
extremizes $S_q$ are (see the Appendix for the definition of the
function involved)
\begin{equation} \label{eq:pi3} p_i^{(3)}=\frac{[1-(1-q)\beta
(\epsilon_{i}-U_q^{(3)}) / \sum_{i=1}^W
{(p_i^{(3)}})^q]^{1/(1-q)}} {\bar Z_q^{(3)}} =\frac{\exp_q[-\beta
(\epsilon_{i}-U_q^{(3)}) / \sum_{i=1}^W {(p_i^{(3)}})^q]} {\bar
Z_q^{(3)}},\end{equation} where
\begin{equation} \label{eq:Z3} \bar Z_q^{(3)}=\sum_{j=1}^W
\left[1-(1-q)\beta (\epsilon_{j}-U_q^{(3)})/ \sum_{i=1}^W
({p_i^{(3)}})^q \right]^{1/(1-q)}  =\sum_{j=1}^W
\exp_q\left[-\beta (\epsilon_{j}-U_q^{(3)})/ \sum_{i=1}^W
({p_i^{(3)}})^q \right].\end{equation} It can be shown that, if
$T\equiv 1/k\beta$ \cite{Tsallis5},
\begin{eqnarray} \frac{1}{T} &=& \frac{\partial S_q}{\partial
U_q^{(3)}},\\
\label{eq:Fq3}F_q^{(3)} &\equiv& U_q^{(3)}-T S_q=-\frac{1}{\beta}
\ln_q  \bar Z_q^{(3)},\end{eqnarray} and
\begin{equation}
S_q=k\ln_q  \bar Z_q^{(3)}.\end{equation} In addition, use of Eq.
(\ref{eq:Fq3}), together with $\beta
\partial U_q^{(3)}/\partial \beta =\partial (\ln_q \bar
Z_q^{(3)})/
\partial \beta$ leads to
\begin{equation}
\label{eq:Uq33} U_q^{(3)}=\frac{\partial}{\partial \beta} (\beta
F_q^{(3)}) \qquad \forall q.\end{equation} Note that  $\bar
Z_q^{(3)}$ refers to the energy levels $\{\epsilon_i\}$, with
respect to $U_q^{(3)}$. We can choose 0 as the energy reference
defining $Z_q^{(3)}$ through
\begin{equation}
\ln_q Z_q^{(3)}=\ln_q \bar Z_q^{(3)}-\beta U_q^{(3)},
\end{equation}
and recast (\ref{eq:Fq3}) and (\ref{eq:Uq33}) as,
\begin{equation}
F_q^{(3)} =-\frac{1}{\beta} \ln_q Z_q^{(3)},
\end{equation}
and
\begin{equation}
U_q^{(3)} = - \frac{\partial \ln_q Z_q^{(3)}}{\partial \beta}.
\end{equation}
Finally, it can be shown that the Legendre structure is preserved
too. The appealing features of this third choice are based in that
it avoids all three problems that were noted for the second set of
possible constraints \cite{Tsallis5}. It is only necessary to
define the new expectation values as
\begin{equation}
O_q^{(3)}\equiv\langle\langle O_i\rangle\rangle_q\equiv
\frac{\sum_{i=1}^W p_i^q O_i}{\sum_{i=1}^W p_i^q},\end{equation}
where $O$ is any observable. Now, $\langle\langle
1\rangle\rangle_q=1 $ $\forall q$ and
\begin{equation}
U_q^{(3)}(A+B)=U_q^{(3)}(A) + U_q^{(3)}(B).\end{equation}

It is important to mention that the probabilities associated with
the third choice coincide with those obtained with the second if
we use a normalized temperature \cite{Tsallis5}. This is the
reason by which all theorems that do not use an explicit
temperature dependence of a given phenomenon will continue to be
valid.

\subsection{Distribution functions}

Despite that the third option is conceptually simpler than the
second, actual computations of thermal dependencies are much
harder. This stems from the fact that in the second option, the
equations for $\{p_i\}$ are explicit (see (\ref{eq:pi2})) whereas
they are implicit in the third case (see (\ref{eq:pi3})). The
objective of this work is to analyze the evolution of the universe
when the distribution functions slightly differ from the standard
ones. To this end, working in the second option will provide the
simplest expressions possible for all observables, what will allow
analytical computations thorough. For particular cases, in
addition, it has been proven that the difference between the exact
results (using the third choice) and the approximation we shall
make here are in very good agreement \cite{TORRES-EPJB}.

Let us now consider a gas composed by $N$ non-interacting
particles. If the system is in thermal contact with a heat and a
particle reservoir, the energy and the number of particles will be
conserved. The stationary state of the system will be given by the
solution of
\begin{equation}
H\psi_R=E_R \psi_R,\end{equation} where  $H$ is the Hamiltonian
and $\psi_R$ is the wave function of the system. The accessible
states will be represented by $R$. In second quantization
formalism, $R$ is given by the set of occupation numbers $\{n_1,
n_2,\ldots, n_k, \ldots\}$, where $n_k$ denotes the number of
particles in the state $k$. Taking $n_k=0,1$ for fermions ensures
Pauli's exclusion principle. We must now extremize the entropy
with the imposed constraints,
\begin{eqnarray} 1&=&\sum_R P_R ,\\ \bar{E}&=&\sum_R P_R^q E_R ,\\
\bar{N}&=&\sum_R P_R^q N_R.\end{eqnarray} To obtain the solution
we use the usual method, extremizing the expression (with units
such that $k=1$),
\begin{equation}
Q=\frac{1}{q-1} \left(1-\sum_R P_R^q \right)-\alpha\sum_R P_R -
\beta \sum_R P_R^q E_R - \gamma \sum_R P_R^q N_R,\end{equation}
where $\alpha, \beta$ are $\gamma$ Lagrange multipliers. Making,
$\partial Q/\partial P_R=0$, we obtain
\begin{equation}
\sum_R \left( \frac{q P_R^{q-1}}{q-1} + \alpha +q \beta P_R^{q-1}
E_R + \gamma q P_R^{q-1}N_R \right)=0.\end{equation} Since it
should be zero for all $R$, each term in the sum must vanish,
\begin{equation}
q P_R^{q-1}+(q-1)\alpha +q (q-1)\beta P_R^{q-1} E_R + \gamma q
(q-1) P_R^{q-1}N_R =0.\end{equation} Identifying $\beta$ and
$\gamma$ by
\begin{equation}
\gamma=-\beta \mu, \qquad \beta=1/T ,\end{equation} where $\mu$ is
the chemical potential and $T$ the temperature, the probability
that the ensamble is in the state $R$ is,
\begin{equation}
\label{eq:PqGC} P_R=[1+\beta(q-1)E_R-\beta(q-1)\mu N_R
]^{1/(q-1)}/Z_q,\end{equation} where
\begin{equation}
\label{eq:ZqGC} Z_q=\sum_R [1+\beta(q-1)E_R-\beta(q-1)\mu N_R
]^{1/(q-1)}.\end{equation} The occupation numbers automatically
determine the quantum state of the system,
\begin{eqnarray} \label{eq:ER}
E_R&=&n_1\epsilon_1+n_2\epsilon_2+\ldots+n_k\epsilon_k+\ldots, \\
\label{eq:NR} N_R&=&n_1+n_2+\ldots+n_k+\ldots, \end{eqnarray}
where  $\epsilon_k$ is the energy of a particle in the state $k$.

Using Eqs. (\ref{eq:ER}) and (\ref{eq:NR}) in the  expressions
(\ref{eq:ZqGC}) and (\ref{eq:PqGC}) we obtain,
\begin{equation}
P_{n_1,\ldots,n_k,\ldots}=\frac{[1+\beta(q-1)(\epsilon_1-\mu)n_1 +
\ldots+\beta(q-1)(\epsilon_k-\mu)n_k +\ldots]^{1/(q-1)}}{Z_q} ,
\end{equation}
with
\begin{equation}
\label{eq:ZqGC2} Z_q=\sum_{n_1,\ldots,n_k,\ldots}
[1+\beta(q-1)(\epsilon_1-\mu)n_1 +\ldots+
\beta(q-1)(\epsilon_k-\mu)n_k + \ldots]^{1/(q-1)}.\end{equation}

The partition function can be factorized as
\begin{equation}
\label{eq:ZqGCfact} Z_q=\prod_{k=1}^\infty \sum_{n_k=0}
[1+\beta(q-1)(\epsilon_k-\mu)n_k]^{1/(q-1)}.\end{equation} It can
be shown that (see  Ref. \cite{TUR} for details) the generalized
occupation numbers are
\begin{equation}
\label{eq:nmq} \langle n_r \rangle_q =
\frac{1}{[1+(q-1)\beta(\epsilon_r-\mu)]^{1/(q-1)}+\xi},\end{equation}
with $\xi=0,+1$ or $-1$ in the case of a Maxwell-Boltzmann (MB), a
Bose-Einstein(BE), or a Fermi-Dirac (FD) gas. In the limit,
$(q-1)\rightarrow 0$ standard distribution functions are
recovered.
\begin{figure} [t]
\vspace{-1.5cm}
\begin{center}
\includegraphics[width=5.5cm,height=8.5cm]{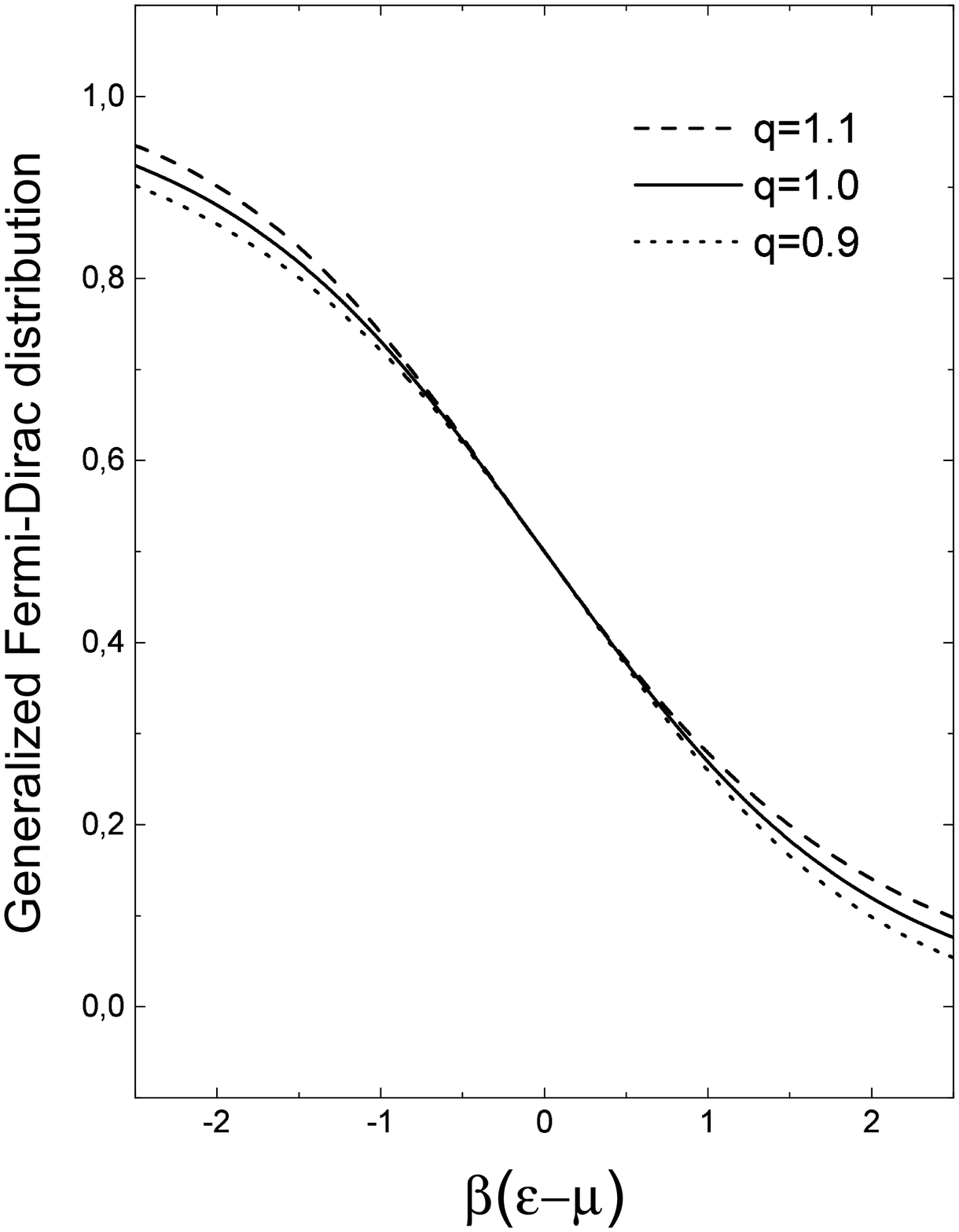}
\includegraphics[width=5.5cm,height=8.5cm]{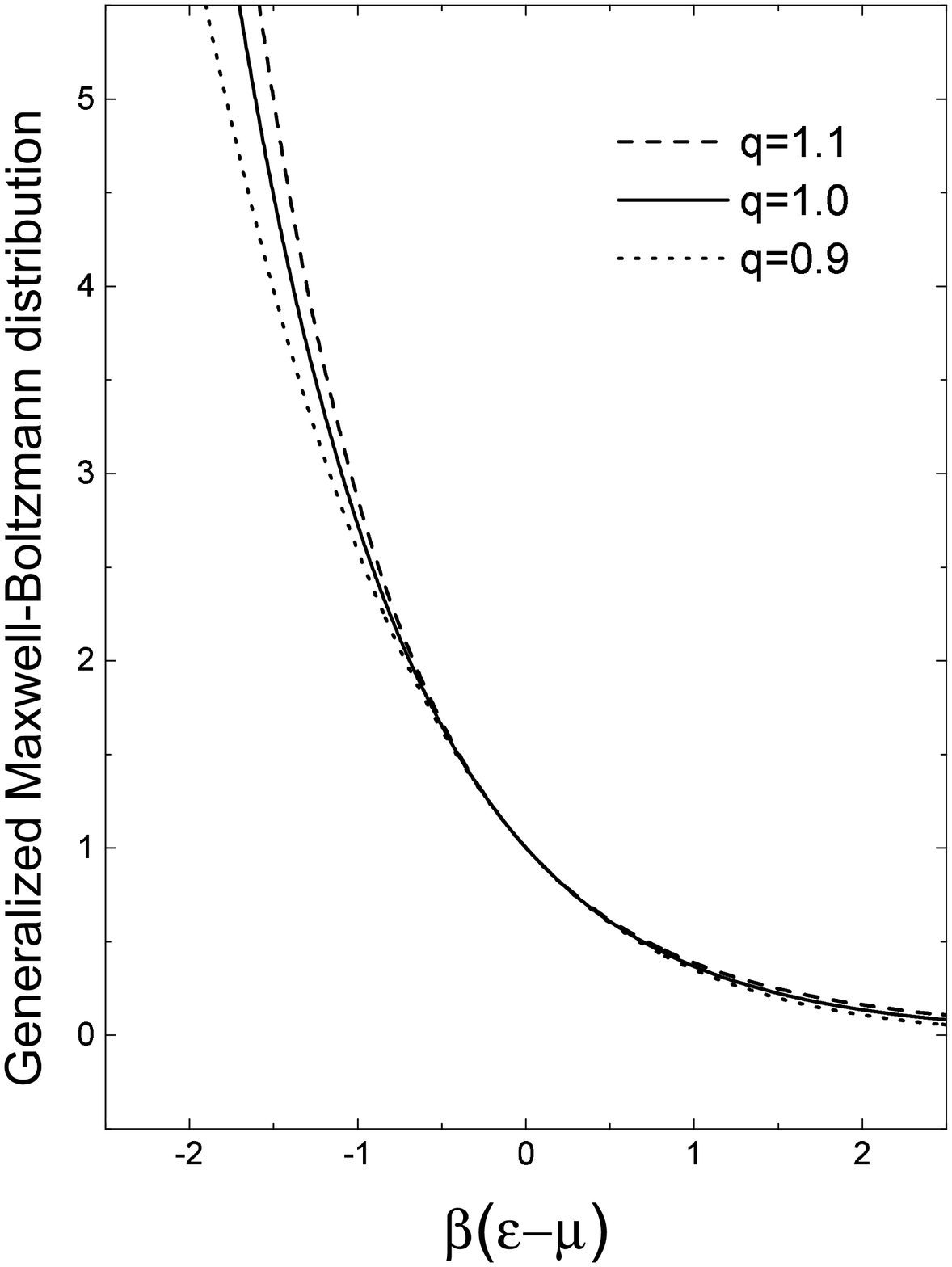}
\includegraphics[width=5.5cm,height=8.5cm]{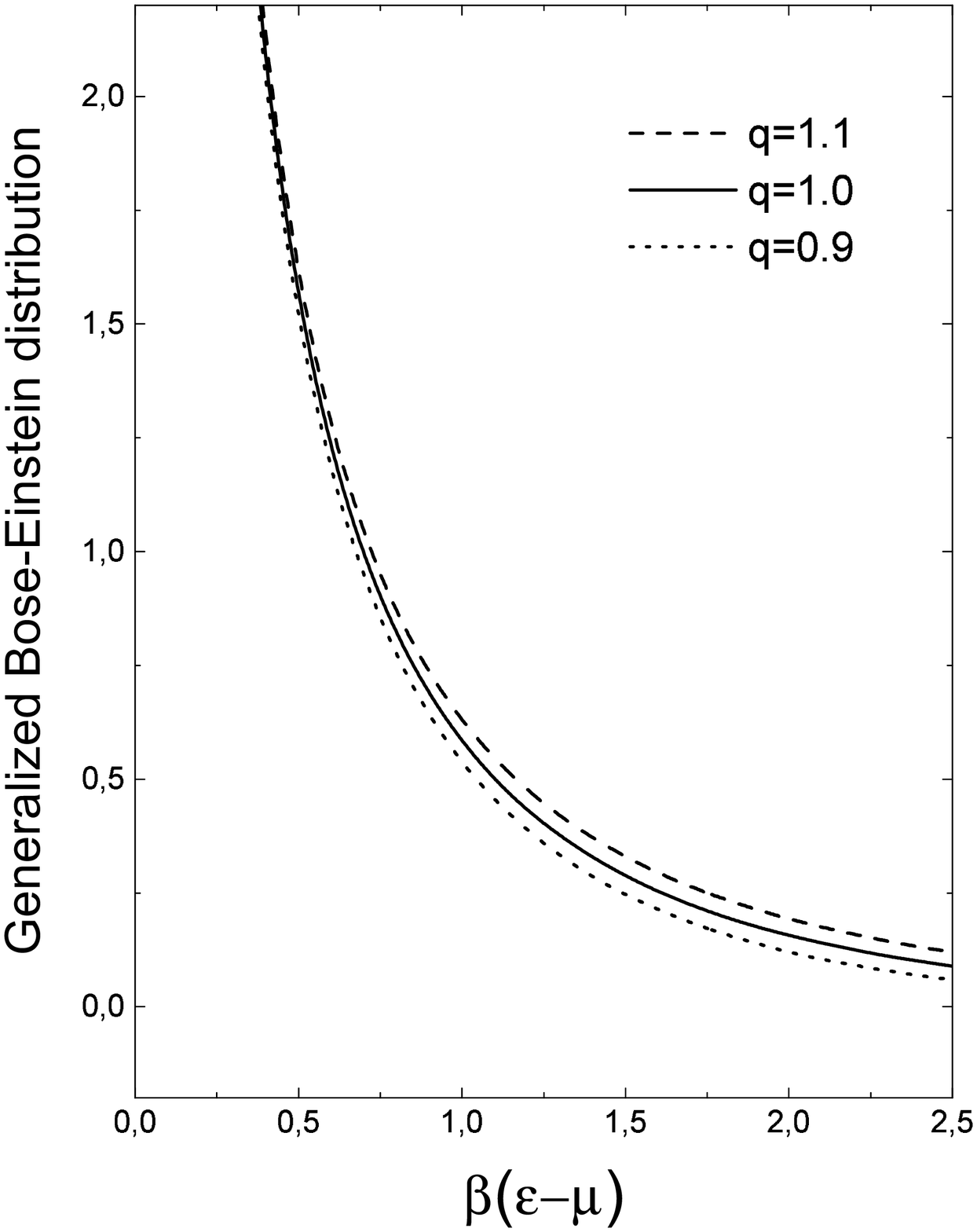}
\end{center}
\vspace{-1.cm} \caption{Behavior of generalized distribution
functions (FD, MB and BE respectively) for $q=1.1$, 1.0 and
$(q-1)=0.9$. } \label{fig:nesdist}
\end{figure}
The expressions for the distribution functions are not exact ones,
since from Eq. (\ref{eq:ZqGC2}) to (\ref{eq:ZqGCfact}) we have
made and approximation, a.k.a. factorization approach \cite{TUR}.
In general, indeed,
\begin{equation}
\label{eq:noteq} [1+(q-1)(A+B)]^{1/(q-1)}\neq
[1+(q-1)A]^{1/(q-1)}[1+(q-1)B]^{1/(q-1)}.\end{equation} Since the
equality is valid for $q=1$ we can certainly wait that for values
$|q-1|\ll 1$ the approximation will remain a good one. We are
interested in the first order in  $(q-1)$, then the expressions
(\ref{eq:nmq}) becomes,
\begin{equation}
\label{eq:nmqfa} \langle n_r \rangle_q =
\frac{1}{e^{\beta(\epsilon_r-\mu)}+\xi}+\frac{q-1}{2}
\frac{(\beta(\epsilon_r-\mu))^2
e^{\beta(\epsilon_r-\mu)}}{(e^{\beta(\epsilon_r-\mu)}+\xi)^2}\end{equation}
In Fig.~1 we show the change in the distribution functions for
different $(q-1)$-values.

Eq. (\ref{eq:nmqfa}) is the expression we were looking for, it is
simple enough, but yet accurate for small deviations, to give an
interesting framework where to analyze (as a parameterization) the
effect of a small change on the extensivity properties of the
universe on its own evolution. Looking at Fig.~1 it can be seen
that despite the particular analytical form for the new
distribution functions, it is clear that they will do an excellent
job in showing an slight deviation from the standard case, this
being the ultimate reason why we use them.

Up to recent days, quantum distribution functions within the
factorization approximation were regarded as a rather rough
technique, see Pennini et al. \cite{pennini}. These latter authors
considered fermion and boson systems with very small occupation
numbers. However, Wang and L\'e M\'ehaut\'e \cite{wang} analyzed
the problem in detail and showed that there exist a temperature
interval, which they called `forbidden zone', where the deviation
from the exact result maybe significant. Otherwise, outside this
zone, the factorization approach results can be used with
confidence. In addition, they verified that the magnitude of the
forbidden zone remained constant with the increase of the number
of particles. This fact motivated new efforts in the study of
macroscopic systems (where the number of particles is large),
since the generalized distribution functions of the factorization
approach could be used at temperatures up to $10^{20}$ K for such
systems \cite{wang}. All these results encourage us further to use
this approximation for the analytical study that follows.

\section{The plasma parameter, the plasma era, and before}

One of the earliest applications of a non-Maxwellian distribution
to plasmas was referred to the solar core. Clayton considered an
ionic quantum distribution function that slightly deviated from a
BG one, being the correction factor proportional to $\exp [-\delta
(E/kT)^2 ]$ \cite{Cla}. $\delta$ can be related with a
non-extensive parameter $q$ by means of the formula $\delta =
(1-q)/2$, see \cite{SOLAR}. There are several reasons why to
expect a slight deviation from a BG behavior in the sun core.
Particularly, we know that long range forces are present (Coulomb
and gravitational) and that there are also effects related with
long range microscopic memory. Then, the leading idea is try to
see whether the solar core is a non-ideal plasma, or a
quasi-plasma, or equivalently, to what extent it can be considered
as ideal. While we forward the reader looking for a detailed
account of these effects to the references quoted in \cite{SOLAR},
let us just mention here the computation of the plasma parameter.

The plasma parameter is the ratio between typical electrostatic
energy of the nearest neighbors of a given particle and the
thermal energy $kT$. For a plasma with number density $n$, the
mean inter-particle distance will be $n^{1/3}$, and it can be
proven that the number of particles inside a Debye sphere is given
by \cite{Padma} \be N_D \sim 360 \left( \frac{T}{{\rm K}}
\right)^{3/2} \left( \frac{n}{{\rm cm}^{-3}} \right)^{-1/2}. \ee
The plasma parameter will be then defined as $\Gamma =
N_d^{-2/3}.$ When $\Gamma \ll 1$, the plasma behaves (just
considering the long range forces) as an ideal gas, the
electrostatic energy is small compared with $kT$ and there are a
large number of particles that shield the long range force in a
Debye-radius range. If, instead, $\Gamma \gg 1$, the plasma is
strongly coupled and certainly non-ideal. For values of $\Gamma$
near 1 we expect slight corrections. For the sun core, for
instance, the plasma parameter is of the order of 0.1 Is it small
enough to produce no appreciable deviation? The answer seems to be
no. It have been recently proven, see last reference in
\cite{SOLAR}, that the effects of random electric microfields are
of crucial importance. These fields have in general long-time and
long-range interactions that can generate anomalous diffusion that
alters the distribution. Most importantly, the amount of the
deviation has been quantified to be proportional to $\Gamma^2$.

The plasma parameter in the early universe is used to define what
is known as the plasma era. At the end of the plasma era is where
the process of recombination occurs. Here, the plasma parameter is
very small and an ideal gas approach is, a priori, a very good
one. If the universe is well described within the standard model
of cosmology, we do not expect the distribution functions to
deviate from the usual BG ones within the plasma era and in later
epochs. Then, if we force the distributions to deviate, the
corrections should turn out to be very big, in such a way that
experiments could then impose restrictive bounds on, for instance,
the non-extensive parameter. This is indeed what happens, as we
shall see below particularly when analyzing the recombination
process. Do the same happens for all eras of cosmic evolution?

Near the temperature of freezing at about 1 MeV, something very
important happens in the universe: the electron-positron
annihilation. At about 0.5 MeV, the cosmological plasma changes
its composition, before the annihilation being formed mainly by
electrons, positrons, photons, neutrinos, and traces of
non-relativistic particles. Which is the value of the plasma
parameter in this era? This plasma is relativistic, and then we
cannot expect that the previously introduced parameter (although
indeed it grows as the temperature of the universe increases) to
be a complete description. We can, however, consider it as a first
approximation. Its value is $\sim 0.07$ at the time of freezing
($t\sim 1$s, $T\sim $1 MeV). This value of $\Gamma$ is close to
the sun core case. Can we a priori expect a similar correction to
the quantum distribution functions? As we shall later show, when
relativistic particles dominate the problem, corrections
introduced by a slight deviation in the distribution functions are
not so big as in the non-relativistic case. Restrictions on the
parameter space are then not so strong, and, a posteriori, it is
not possible to straightforwardly discard such slight deviations.
In addition, if $\Gamma^2$ gives the amount of the deviation also
for this system, we would expect corrections of the order of
$10^{-2}-10^{-3}$, what we do obtain (independently of these
comments) when we analyze the epoch of primordial nucleosynthesis,
see paper II in this series \cite{P2}.

We do not claim here that these paragraphs provide evidence of
such an strong nature as to consider a priori that deviations to
BG distributions functions are unavoidable, as the analysis for
the sun core case seem to show \cite{SOLAR}. We do believe,
however, that they make worth looking at the standard model
predictions in an statistically modified setting, to see the
allowed range of the deviations along the different cosmic eras.
To that aim we devote this and the follow up paper in the series.
On the other hand, we recall that several authors, with various
motivations, have considered other effects that could slightly
change the distribution functions as well: Non-equilibrium effects
originating in residual interactions between neutrinos and
electrons (see for instance \cite{Dolgov}) and fluctuations in
magnetic fields possibly present at that time (see for instance
\cite{Opher}) are some examples.

\section{Thermodynamics in the expanding universe}

\subsection{Energy density, pressure, number of particles}

The number density, the energy density, and the pressure for a
dilute gas are given in terms of their corresponding distribution
function $f({\bf p})$ by (units are such that $\hbar=c=1$),
\be
\label{number} n=\frac {g}{(2\pi)^3} \int f({\bf p}) d^3p ,
\ee
\be
\label{energy} \rho=\frac {g}{(2\pi)^3} \int f({\bf p})
E(|{\bf p}|) d^3p ,
\ee
\be
\label{p} P=\frac {g}{(2\pi)^3} c^2 \int
f({\bf p}) \frac { | {\bf p } |^2}{3E(|{\bf p}|)} d^3p ,
\ee
where $E(|{\bf   p}|)=| {\bf p } |^2 + m^2$, and $g$ is the degeneracy
factor of the particle. From now on, $E(|{\bf   p}|)=E(p)$ and $|
{\bf p } |=p$. Let us first take the non-degenerate case, in which
$kT \gg \mu$. As before, $T$ will stand for the temperature and
$\mu$ for the chemical potential. Use of the form of $f(p)$ in the
non-extensive approach we adopted gives the correction to the
standard result at first order in $(q-1)$. Several cases are worth
mentioning.

When the particles are relativistic, $E \gg m$, the energy density
is
\be \rho_q=\rho_{st} +\frac 1{2\pi^2} \frac {5!}{2} (1.04
g_{bosons} + 0.97 g_{fermions}) (q-1)  T^4 ,
\ee with $\rho_{st} $
representing the usual result, \be \rho_{st} =  \frac {\pi^2}{30}
g T^4 \ee and $g=  g_{bosons} + 7/8 g_{fermions}$. For the number
of particles, we obtain, \be n_q^{bosons}= \frac {g}{2\pi^2}2
\zeta(3) T^3 +  \frac {g} {2\pi^2} 12.98 (q-1) T^3,
\label{eq:nqb}\ee \be n_q^{fermions}= \frac {g}{2\pi^2}2 \frac 32
\zeta(3) T^3 +  \frac {g} {2\pi^2} 11.36 (q-1) T^3.
\label{eq:nqf}\ee For the pressure, starting from the definition,
we can immediately prove that $P=\rho/3$
disregarding the degree of non-extensivity. \\

{\it Remark 1: Note that, for relativistic particles, the
temperature dependence of $\rho, P$ and $n$ in the non-extensive
formalism is the same as the standard. This is because it comes
only as a product of the change of variables in the integrals,
which is done exactly in the same way for both contributions.}
\\

It is interesting now to compute the energy per particle, which results to
be
\be
<E_q>^{bosons}=[2.70 + 11.29 (q-1)] T, \ee
\be
<E_q>^{fermions}=[3.15 + 44.83 (q-1)] T. \ee Note that to equal
$q$, the fermions get more affected than the bosons by the change
of the distribution.

In the non-relativistic limit $m \gg T$, we can neglect the $\pm
1$ term in the denominator of all integrals. For instance, the
number density becomes,

\be
\label{nr} n_q=g \left(\frac{mT}{2\pi}\right)^{3/2} e^{-(m-\mu)/T}
\times \left[ 1 + \frac{q-1}{2} \left( \frac {15}4 + 3 \frac
{m-\mu}{T} + \left(\frac {m-\mu}{T}\right)^2 \right) \right]\ee
Here, the first term in the right hand side stands for the
standard result and the second three for the correction at $(q-1)$
order. The energy density is then given by $\rho_q = mn_q$ and the
pressure is $P_q = n_q T$.
\\

{\it Remark 2: For non-relativistic particles, the change to a
non-extensive setting does change the temperature dependence of
the observables. This is not a minor effect and will show its
impact in the recombination process.}

\subsection{Particle anti-particle excess}

We focus now on the computation of the particle anti-particle
excess. Consider the reaction $r^+ + r^- \leftrightarrow \gamma +
\gamma$, where $r^+$ and $r^-$ represent a generic particle and
its anti-particle. If this reaction is in equilibrium, then the
chemical potential of $r^+$ and $r^-$ are related by $\mu^+ = -
\mu^-$. Then, the net number of fermions is given by \be n^+_q -
n^-_q = \frac{g}{2\pi^2} T^3 \int_{m/T}^\infty x\left(x^2- \left(
\frac{m}{T}\right)^2 \right)^{1/2} \left[ f_q^+ (x) -
f_q^-(x)\right] dx, \ee where the integration variable is, as
before,  $x=E/T$. In the relativistic limit, $m/T \ll 1$, and the
standard integrals can be solved as \be n_{st}^+ - n_{st}^-=
\frac{g}{6\pi^2} T^3 \left[ \pi^2 \frac{\mu}{T} + \left(
\frac{\mu}{T}\right)^3\right]. \ee To compute the $(q-1)$ order
corrections we need to calculate \be \label{excess}
\frac{g}{2\pi^2} T^3 \frac{q-1}{2}\int_{0}^\infty \left[ \frac{x^2
(x-\psi)^2 e^{x-\psi}}{(e^{x-\psi} + 1)^2 } - \frac{x^2 (x+\psi)^2
e^{x+\psi}}{(e^{x+\psi} + 1)^2 } \right] dx. \ee Since we would
like to have an analytical solution, we think in a new application
of the free parameter trick introduced in this context by
T{\i}rnakl{\i} and Torres \cite{TORRES-EPJB}. We know that \be
I^{(m)}=\int_0^\infty \frac {u^2}{e^{m(u-\psi)}+1}- \int_0^\infty
\frac {u^2}{e^{m(u+\psi)}+1}= \frac 13 \left(\pi^2 \psi m^{-2} +
\psi^3 \right). \ee Here, $m$ is considered as a free parameter.
Then, the integral we need can be obtained by making the following
steps. Consider $ - d^2I^m/ dm^2$, evaluated at $m=1$. This last
integral will give the one we want plus two additional terms.
These terms are \be \int_0^\infty du \left[\frac {-2 u^2
(u-\psi)^2 e^{m(u-\psi)}}{(e^{m(u-\psi)}+1)^3}- \frac {-2 u^2
(u+\psi)^2 e^{m(u+\psi)}}{(e^{m(u+\psi)}+1)^3} \right]. \ee We can
evaluate the relative strength of these extra terms with respect
to the integrals we are looking for. The first thing one can
analytically do is to consider the case in which $\psi \sim 0$.
This, we know, is not far from the real situation, since after
electron positron annihilation, only a little excess of electrons
will survive, in order to give the universe, together with
protons, electrical neutrality. Because the ratio of the number
density of protons to the number density of photons,
$n^p_{st}/n_\gamma^{st} \sim 10^{-9}$, use of the standard
expression for $n^+_{st} - n^ -_{st}$ and $n_\gamma^{st}$ will
show that \be \frac{n^+ - n^-}{n_\gamma} \sim 1.33 \frac {\mu}{T}
\sim 10^{-9}, \ee and clearly the condition $\mu/T \ll 1$ is
sustained. Taking this into account we can weight each of these
two extra terms with respect to our integrals. These terms are,
\be A = \int_0^{\infty} \frac{u^4 e^u}{(e^u +1)^2} du, \ee and \be
B = 2 \int_0^{\infty} \frac{u^4 e^u}{(e^u +1)^3} du. \ee If we
call $f_A(u)$ and $f_B(u)$ to the integrands of the previous
expressions, we note that \be f_B(u) = \frac{2}{e^u + 1} f_A(u).
\ee Around $0$, $f_B(u) \simeq f_A(u)$; but the integrand $B$
falls exponentially. Finally, a numerical integration of the two
terms, that we show in Fig.~1, yields a difference bigger than one
order of magnitude: $A = 22.63$ and $B= 1.037$. For the sake of
analytical computation, we shall adopt the following approximation
for the correction to the particle anti-particle excess $n_c$, \be
n_c^+ -n_c^- \simeq \frac {g}{2\pi^2} T^3 \frac{q-1}{2} \left[
\frac{ -d^2I^m}{dm^2}\right]_{m=1}, \ee which finally yields the
result, \be n^+_q -n^-_q = \frac {g}{6\pi^2} T^3 \left[ \pi^2
\left( 1+3(1-q)\right) \frac{\mu}{T} + \left( \frac{\mu}{T}
\right)^3 \right]. \ee

\begin{figure}
\leavevmode \epsfxsize=8.5cm \epsfysize=9.5cm \centering
\epsffile{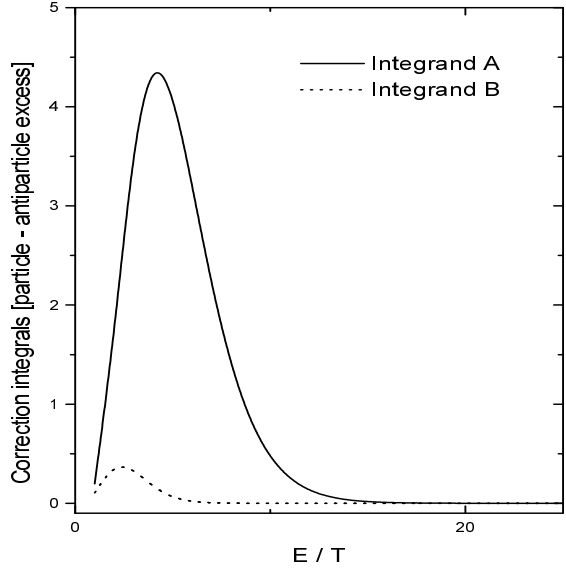} \vspace{-1cm}\caption{Relative strength of
the integrals appearing in the computation of $n^+_c - n^-_c$. }
\label{Fig1}
\end{figure}

In the case of non-relativistic particles, a similar computation
gives \be n^+_q -n^-_q = (n_{st}^+ -n_{st}^-) + (n_c^+ -n_c^-),
\ee with the standard term being \be n_{st}^+ -n_{st}^- = g
\left(\frac {mT}{2\pi} \right)^{3/2} e^{-m/T} 2 \sinh (\mu/T), \ee
and the correction being \be n_c^+ -n_c^- =  2g \left(
\frac{mT}{2\pi} \right) ^{3/2} e^{-m/T} \left[ \sinh \left(
\frac{\mu}{T} \right) \left[ \frac{15}{4} + 3 \frac mT  + \left(
\frac{m^2+\mu^2}{T^2} \right) \right ] - \cosh \left(
\frac{\mu}{T} \right) \left[ 3 \frac{\mu}{T} + 2\frac{m\mu}{T}
\right] \right]. \ee

\subsection{Effective number of degrees of freedom}

The total energy density of all species in equilibrium can be
expressed in terms of the photon temperature. As most of the
energy density is contributed by relativistic species, we shall
firstly consider only them. The total density will be the sum over
all particular species, \be \rho_R= \sum_k \rho_k= \sum_{bosons}
\rho_i + \sum_{fermions} \rho_j . \ee Note that we are using a
sub-index $i$ for boson particles, whereas $j$ is used for
fermions. From previous expressions we have, taking $x_r=E_r/T$,
\be \rho_i= \frac{g_r T_r^4}{2\pi^2} \int_{m_r/T_r}^\infty x_q^3
f_q^r(x_r) dx_r, \ee recall here that $f_q^r(x_r)$ is the
non-extensive distribution function, and we are in the limit in
which $E_r \gg m_r$ and $x_r \gg m_r/T_r$ are valid. If we as well
consider $T_r \gg \mu_r$, then $f_q^r(x_r)\equiv f_q^r(x)$. This
gives the separate corrections, \be \rho_{bosons}=\rho_{bosons, \;
st} + \frac {g_i}{2\pi^2} \frac {q-1}{2} 5! \zeta(5) T_i^4 ,\ee
\be \rho_{fermions}=\rho_{fermions, \; st} + \frac {g_j}{2\pi^2}
\frac {q-1}{2} 5! f_5(1) T_j^4 .\ee The standard contributions are
the usual, \be \rho_{bosons, \; st} = \frac {g_i \pi^2}{30}
T_i^4,\;\;\;\;\; \rho_{fermions, \; st} = \frac 78 \frac {g_j
\pi^2}{30} T_j^4. \ee Adding up both contributions, we can write
\be \label{rel} \rho_R = \frac{\pi^2}{30} g_*^q T^4, \ee where
$g_*^q$ is given by the addition of the standard, \be \label{a}
g_*^{st}= \sum_{bosons} g_i \left(\frac{T_i}{T}\right)^4 + \frac
78 \sum_{fermions}
 g_j \left(\frac{T_j}{T}\right)^4
\ee and the new correction, \be \label{b} g_*^c= (q-1) \left[ 9.58
\sum_{bosons} g_i \left(\frac{T_i}{T}\right)^4 + 8.98
\sum_{fermions}
 g_j \left(\frac{T_j}{T}\right)^4 \right].
\ee

{\it Remark 3: Note that, due to the same temperature dependence,
we can define an effective number of degrees of freedom exactly in
the same way as it is done in the standard case. Then, we can
consider the same evolution equations, and hide all
non-extensivity effects in the new $g_*^q$.}\\

As an example of the previous remark, we can consider the
Friedmann equation. It reads, \be \frac {\dot R}{R} + \frac k{R^2}
= \frac{8\pi G}{3} \rho, \ee where $R(t)$ is the scale factor in
the FRW metric, and $k$ is the curvature of the space-time. In the
early universe, it is a very good approximation to take $k=0$.
During the radiation dominated phase $P=\rho/3$, $R(t)=t^{1/2}$,
and $ \dot R /R=H(t)=1/t$. Using the expression for the energy
density, i.e. Eq. (\ref{rel}), we obtain \be H =1.66 \left( g_*^q
\right)^{1/2} \frac{T^2}{m_{Pl}} , \ee implying, \be t=0.30 \left(
g_*^q \right)^{-1/2} \frac {m_{Pl}}{T^2}. \label{eq:tTrel}\ee
Then, although the functional form is the same as in the standard
case, the actual time-temperature relationship is different
because of the change in the quantum distributions.

\section{Conserved quantities}
\mbox{}From the full set of Einstein field equations, or from the
fact that $T^{\mu\nu}_{\;\;\;\;;\nu}=0$, it may be established
that

\be d(\rho_q R^3) = -P_q d(R^3). \ee This can be written as \be
\label{rel1} \frac{d}{dT} \left[\left( \rho_q + P_q
\right)R^3\right] = R^3 \frac{dP_q}{dT}. \ee And using the
previous equation we finally get \be \label{rel2} \frac{d}{dT}
\left[ \frac{\rho_q + P_q }{T} R^3\right] = 0, \ee as can be
checked by direct differentiation.
\\

{\it Remark 4: Disregarding the degree of non-extensivity, $s_q
\equiv (\rho_q + P_q)/ T$ is a conserved quantity in a comoving
volume.}\\

If the relativistic contribution is dominant, \be s_q = \frac 43
\left[ \sum_{bosons} \frac{\rho_i}{T_i} + \sum_{fermions}
\frac{\rho_j}{T_j} \right] . \ee This yields \be s_q =
\frac{2\pi^2}{45} g_{*,s}^q T^3, \ee where we have defined
$g_{*,s}^q$ as \be \label{s} g_{*,s}^q = g_{*,s}^{st} + g_{*,s}^c,
\ee with \be g_{*,st}^s= \left[ \sum_{bosons} g_i
\left(\frac{T_i}{T}\right)^3 + \frac 78 \sum_{fermions} g_j
\left(\frac{T_j}{T}\right)^3 \right] ,  \label{eq:gsst} \ee \be
\label{eq:gsc} g_{*,s}^c=7.18 (q-1) \left[ \sum_{bosons} g_i
\left(\frac{T_i}{T}\right)^3 + \frac {15}{16} \sum_{fermions} g_j
\left(\frac{T_j}{T}\right)^3 \right] .\ee From the conserved
comoving quantity, we may define the number $S_q=s_q R^3 =
constant$, and obtain the temperature-scale factor relationship as
\be T \propto (g_{*,s}^q)^{-1/3} R^{-1}. \ee Then, as in the
standard case, $T \propto R^{-1}$, but now only in the periods
where $g_{*,s}^q$ (and not $g_{*,s}$) is constant.

\subsection{Explicit form for the conserved numbers}

For reasons that clarify themselves when actually making
computations, it is convenient to define $\hat g_{\star s}^c$ such
that $ g_{\star s}^c=(q-1) \hat g _{\star s}$. As in the standard
case, we have proven that for any particle which is not created
nor destroyed during the evolution of the universe, \be N_q =\frac
{n_q} {s_q} \ee is a conserved quantity. For future use, we quote
here the explicit form of the conserved number for relativistic
particles. For instance, for photons, the number density is
$n^\gamma_q  = g/(2\pi^2) \left[ 2\zeta(3) + 12 \zeta(4) (q-1)
\right] T^3$, with $g=2$. Then, \be s_q= \frac {2\pi^4}{45} \left[
2\zeta(3) + 12 \zeta(4) (q-1) \right]^{-1} g_{*,s}^q n^\gamma_q .
\ee

This gives,
\be
N_q^b=N_{st}^b + \frac{45}{2\pi^4} g^b (q-1) \left[ 6\zeta(4)-
\frac{\hat g_{\star s}^c}{g_{\star s}^{st}} \zeta(3) \right], \ee
with
\be
N_{st}^b=\frac{45}{2\pi^4} \zeta(3)g^b \left( g_{\star s}^{st}
\right)^{-1}. \ee

Similar computations, but now for fermions, would yield, \be
N_q^f=N_{st}^f + \frac{45}{2\pi^4} \frac{3}{4} g^f (q-1) \left[ 21
\zeta(4)- \frac{\hat g_{\star s}^c}{g_{\star s}^{st}} \zeta(3)
\right], \ee with \be N_{st}^f=\frac{45}{2\pi^4} \frac{3}{4}
\zeta(3) g^f \left( g_{\star s}^{st} \right)^{-1}.\ee

\section{Decoupling}

Once a given species of particles has decoupled completely, each
particle will travel along a geodesic of space-time. This will
ensure that the form of the distribution function will remain the
same during the subsequent evolution of the universe. Indeed, the
only change will be given by a redshift correction. If the
decoupling is produced at $t =t_d$, then, due to the redshift, all
particles that at the instant $t$ have momentum ${\bf p}$, should
have had momentum ${\bf p} R(t)/R(t_d)$ at decoupling. Then, \be
\label{eq:dec=eq} f_{dec}({\bf p},t)=f_{eq}({\bf
p}R(t)/R(t_d),t_d),\;\;\forall t>t_d. \ee

{\it Remark 5: The previous result is independent of the particular
$f_{eq}({\bf p},t) $ we have adopted.}\\

\subsection{Relativistic decoupling}

Let us write the rhs of Eq. (\ref{eq:dec=eq}) taken into account
that $E \simeq p$ and in the case in which $\mu/T \simeq 0$, \be
\label{eq:fqdrnd} f_q^d({\bf p},t) =
\left[\exp{\left(\frac{p}{T_d}\frac{R(t)}{R(t_d)}\right)}\mp1\right]^{-1}
+\frac{q-1}{2} \frac{ \left( \frac{p}{T_d}\frac{R(t)}{R(t_d)}
\right)^2 \exp{\left(\frac{p}{T_d} \frac{R(t)}{R(t_d)}
\right)}}{\left[\exp{\left(\frac{p}{T_d}\frac{R(t)}{R(t_d)}\right)}\mp1\right]^2}.
\ee We see that the distribution function has the same functional
form than $f_q^{eq}$ but with a different ``temperature'' given
by, \be T(t)=T_d \frac{R(t_d)}{R(t)}; \ee Despite this species is
no longer in thermodynamical equilibrium, the ``temperature'' in
the distribution function falls as $R^{-1}$; making $S_q^r=s_q^r
R^3$ to conserve separately. Note that for the species still in
equilibrium, the temperature falls as  $T \propto (g_{\star
s}^q(T))^{-1/3} R^{-1}$. The number density for these particles is
then given by, \be n_q^{b_r}=\frac{g^{b_r}}{2\pi^2} \left[
2\zeta(3) + 12 \zeta(4) (q-1) \right] T^3, \ee \be
n_q^{f_r}=\frac{g^{f_r}}{2\pi^2} \left[ \frac{3}{2}\zeta(3) +
\frac{21}{2} \zeta(4)(q-1) \right] T^3. \ee and using that for any
time $t$, after $t_d$, $T(t)=T_d \frac{R(t_d)}{R(t)}$, the
previous equations really are, \be
n_q^{b_r}=\frac{g^{b_r}}{2\pi^2} \left[ 2\zeta(3) + 12 \zeta(4)
(q-1) \right] T_d^3 \left(\frac{R(t_d)}{R(t)} \right)^3, \ee \be
n_q^{f_r}=\frac{g^{f_r}}{2\pi^2} \left[ \frac{3}{2}\zeta(3) +
\frac{21}{2} \zeta(4)(q-1) \right] T_d^3 \left(\frac{R(t_d)}{R(t)}
\right)^3. \ee These number densities are comparable to that of
photons at any given time. Such population will then continue to
exist as a relativistic relic.

\subsection{Non-relativistic decoupling}

In the case of non-relativistic decoupling, the distribution
function is given by \begin{equation} f_q^d({\bf p},T)=
e^{-(m-\mu_d)/T_d}\exp{\left [\frac{-p^2}{2mT_d} \left(
\frac{R(t)}{R(t_d)} \right)^2 \right]} \left[ 1 + \frac{q-1}{2}
{\left [\frac{p^2}{2mT_d}\left( \frac{R(t)}{R(t_d)} \right)^2
+\frac{(m-\mu_d)}{T_d} \right]}^2 \right], \end{equation} and the
number density is given by
\begin{equation}
n_q=g \left( \frac{mT_d}{2\pi} \right)^{3/2} \left(\frac
{R(t_d)}{R(t)}\right)^3 e^{-(m-\mu_d)/T_d} \left[1 + \frac{q-1}{2}
\left(\frac{15}{4}+3\frac{m-\mu_d}{T_d}+
\left(\frac{m-\mu_d}{T_d}\right)^2\right)\right]. \end{equation}
If we now consider that for a particle species which has already
decoupled, the number density must be such that $n_q\propto
R^{-3}$, we see that as in the standard case (see for instance
Section 3.4 of Ref.\cite{Kolb}), we need to identify the
temperature with $T_d R^2(t_d)/R^2(t)$, and ask for the chemical
potential to vary as $\mu=m-(m-\mu_d)T/T_d$. Then, the
distribution function is similar to an equilibrium distribution
but with a new temperature given by \be T(t)=T_d
\left(\frac{R(t_d)}{R(t)}\right)^2, \ee decreasing with the square
of the scale factor. In the limit $m \gg T$, the energy
density will be given by $\rho = n m $.\\

Also in a non-extensive framework, the correct way to handle with
decoupling process is through the Boltzmann equation, which we
analyze below.

\subsection{The relic neutrino background}

The conservation of $s_q$, applied to the particles which are in
equilibrium with radiation shows that the quantity $g_{*,s}^q
T_\gamma^3 R^3$ is constant throughout the expansion. During pair
annihilation, $g_{*,s}^q$ will decrease, since we go towards a
state with less number of degrees of freedom. Then, $T_\gamma^3
R^3$ after the process will be bigger than its previous value. We
say that photons are heated up by the annihilation process, in
such a way that \be (g_{*,s}^q T_\gamma^3 R^3
)_{before}=(g_{*,s}^q T_\gamma^3 R^3 )_{after} \ee is sustained.
This directly yields a relationship between both temperatures, \be
\frac {T_\gamma^{after}}{ T_{\gamma}^{before}} = \left( \frac
{11}{4} \right)^{1/3} + 0.152 (q-1)=  \left( \frac {11}{4}
\right)^{1/3} (1+ 0.109 (q-1))\ee Note that the first term is the
standard result, from the usual hot big bang model. The second
term is the non-extensive correction. \\

Since before $e^+ e^-$ annihilation, neutrinos were in thermal
equilibrium with photons, $T_\gamma ^{before}$ will be the same as
$T_\nu ^{before}$. Now notice that neutrinos do not participate in
the process $e^+ + e^- \leftrightarrow 2 \gamma$, and so, neutrino
temperature is constant. Then, $T_\nu ^{before}=T_\nu ^{after}$,
and we can write \be T_\gamma^{after}= \left[ \left( \frac {11}{4}
\right)^{1/3} + 0.152 (q-1)  \right] T_\nu ^{after}.
\label{eq:Tnu/Tg-hoy}\ee After annihilation, $g_{*,s}^q $ does not
change anymore, and both temperatures $T_\gamma^{after}$ and
$T_\nu ^{after}$ fall as $R^{-1}$, conserving their ratio. Since
now $T_\gamma \sim 2.728$, then
$T_\nu=1.947 (1- 0.109(q-1))$K.\\

{\it Remark 6: Even if the evolution of the universe diminish the
degree of non-extensivity, the relic backgrounds, produced at
early times, are sensitive to the possible non-extensivity
present at that epoch.}\\

{\it Remark 7: Note that if $(q-1)$ could have any value, one can
have a photon temperature unchanged after $e^+ e^-$ annihilation:
$T_\gamma^{after}/T_{\gamma}^{before} =1$. This effect
particularly show how important is the statistical description in
the evolution of the early universe. Note, however, that a
posteriori constraints on the value of $(q-1)$ do apply, and this
situation will turn out to be
impossible.}\\

The difference between $T_\nu$ and $T_\gamma$ can be bounded using
information coming from the CMBR, see for instance the work by
Torres \cite{TORRES-PHYSA}.

\section{Recombination}

We focus now on the generalization of the Saha law. Let $n^H,n^p$
and $n^e$ the number density of hydrogen atoms, protons, and
electrons respectively. Electrical neutrality implies that
$n^e=n^p$. Baryon total number is given by $n^B=n^H+n^p$. In
thermal equilibrium, a temperatures such that $T < m_i$, we have
seen that $n_i$ is given by Eq. (\ref{nr}). The process $p+e
\rightarrow H+p$ guarantees that $\mu_p+\mu_e=\mu_H$. We can as
well consider that $m_p \simeq m_H$. Then, we would like to find
an expression for $n^H_q/n^e_q n^p_q$. Immediate algebra yields,
\be \frac{n^H_q}{n^e_q n^p_q}= \frac {g_H}{g_e g_p} \left( \frac
{m_e T}{2 \pi} \right)^{-3/2} {\rm exp}\left[ \frac{m_p +m_e
-m_H}{T} \right] \left[\frac{\bar u((m_H-\mu_H)/T)}{\bar
u((m_p-\mu_p)/T)\bar u((m_e-\mu_e)/T)}\right]. \ee Here, we have
made use of the definition, \be \label{ubar}\bar u_i= \bar
u((m_i-\mu_i)/T)\equiv 1 + \frac{q-1}{2} \left( \frac {15}4 + 3
\frac {m_i-\mu_i}{T} + \left(\frac {m_i-\mu_i}{T}\right)^2
\right). \ee We would like to introduce now the number of baryons,
since this is a conserved number (as far as precision measurements
can tell), i.e. $N^B=n^B/s$ is constant. Note however that
$\eta_q=n^B_q/n^\gamma_q$ does not remain unchanged, since
$g_{*,s}^q$ may change in time. Using $g_p=g_e=2$ and $g_H=4$ we
can write, \be \frac{n^H_q}{n^B_q}=\frac{n^p_q}{n^B_q}
\frac{n^e_q}{n^B_q} n_\gamma^q \eta_q \left( \frac {m_e T}{2 \pi}
\right)^{-3/2} {\rm exp}\left[ \frac{m_p +m_e -m_H}{T} \right]
\left[\frac{\bar u((m_H-\mu_H)/T)}{\bar u((m_p-\mu_p)/T)\bar
u((m_e-\mu_e)/T)}\right]. \ee We now define $X^e_q=n^p_q/n^B_q$
and $1-X^e_q=n^H_q/n^B_q$. Using them, we have \be
\frac{1-X^e_q}{(X^e_q)^2}=n^\gamma_{st}
\left[1+\frac{n_{\gamma,c}}{n^\gamma_{st}}\right] \eta_q \left(
\frac {m_e T}{2 \pi} \right)^{-3/2} e^{B/T} \left[\frac{\bar
u_H}{\bar u_p\bar u_e}\right], \ee where $B=m_p +m_e -m_H$ and
$n_\gamma^q=n_{\gamma,st}+(q-1)n_{\gamma,c}$. Note that we have
shortened the notation on the last bracket, but we are actually
implying the definition of Eq. (\ref{ubar}). Also, \be
\eta_q=\eta_{st} \left[1+\frac{n^B_c}{n^B_{st}}\right]\left[
1+\frac{n^\gamma_c}{n^\gamma_{st}} \right]^{-1}. \ee Finally, we
rewrite \be \left[\frac{\bar u_H}{\bar u_p\bar u_e}\right] = 1+
\frac{q-1}{2} \left[ u_H - (u_p + u_e)\right], \ee with \be u_i=
\left( \frac {15}4 + 3 \frac {m_i-\mu_i}{T} + \left(\frac
{m_i-\mu_i}{T}\right)^2 \right), \ee and we use that  \be
\frac{n^B_c}{n^B_{st}}=\frac{n^H_c+n^p_c}{n^H_{st}+n^p_{st}}=
\frac{q-1}{2} \left[ \frac{g_H e^{(\mu_H - m_H)/T} u_H +g_p
e^{(\mu_p - m_p)/T} u_p} {g_H e^{(\mu_H - m_H)/T} + g_p e^{(\mu_p
- m_p)/T}} \right]. \ee Recalling that $n^\gamma_{st}= (2
\zeta(3)/\pi^2) T^3$, we can write, with $g_H=$, and $g_p=g_e=2$,
\ben \frac{1-X^e_q}{(X^e_q)^2}= \hspace{15cm}\nonumber \\ \frac
{4\sqrt{2}}{\sqrt{\pi}} \zeta(3) \eta_{st} \left( \frac {T}{m_e}
\right)^{3/2} e^{B/T} \left[ 1 + \frac{q-1}{2} \left( \frac{2
e^{(\mu_H - m_H)/T} u_H + e^{(\mu_p - m_p)/T} u_p} {2 e^{(\mu_H -
m_H)/T} +  e^{(\mu_p - m_p)/T}} + u_H - (u_e+u_p) \frac{}{}\right)
\right]. \een And taking into account the standard result, \be
\frac{1-X^e_{st}}{(X^e_{st})^2}= \frac {4\sqrt{2}}{\sqrt{\pi}}
\zeta(3) \eta_{st} \left( \frac {T}{m_e} \right)^{3/2} e^{B/T} \ee
we see that we have arrived to the following relationship: \ben
\label{gsl} \frac{1-X^e_q}{(X^e_q)^2}= \hspace{15cm}\nonumber \\
\frac{1-X^e_{st}}{(X^e_{st})^2}+ \frac{q-1}{2} \frac
{4\sqrt{2}}{\sqrt{\pi}} \zeta(3) \eta_{st} \left( \frac {T}{m_e}
\right)^{3/2} e^{B/T} \left( \frac{2 e^{(\mu_H - m_H)/T} u_H +
e^{(\mu_p - m_p)/T} u_p} {2 e^{(\mu_H - m_H)/T} +  e^{(\mu_p -
m_p)/T}} + u_H - (u_e+u_p) \frac{}{}\right). \een This is the
generalized Saha law.

\begin{figure}[t]
\centering \leavevmode \epsfxsize=8.5cm \epsfysize=9.5cm \label{f}
\epsffile{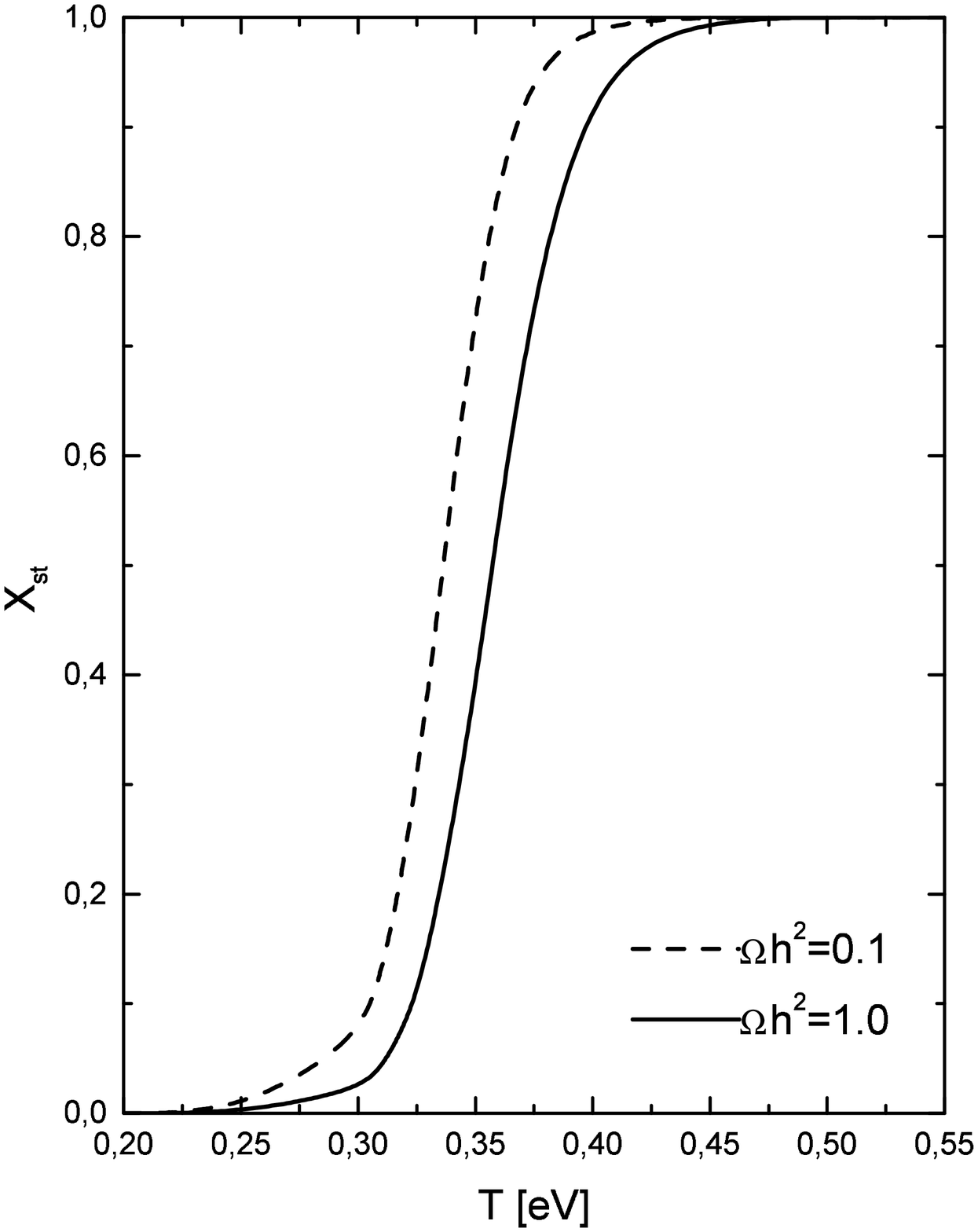} \epsfxsize=8.5cm \epsfysize=9.5cm \label{m}
\epsffile{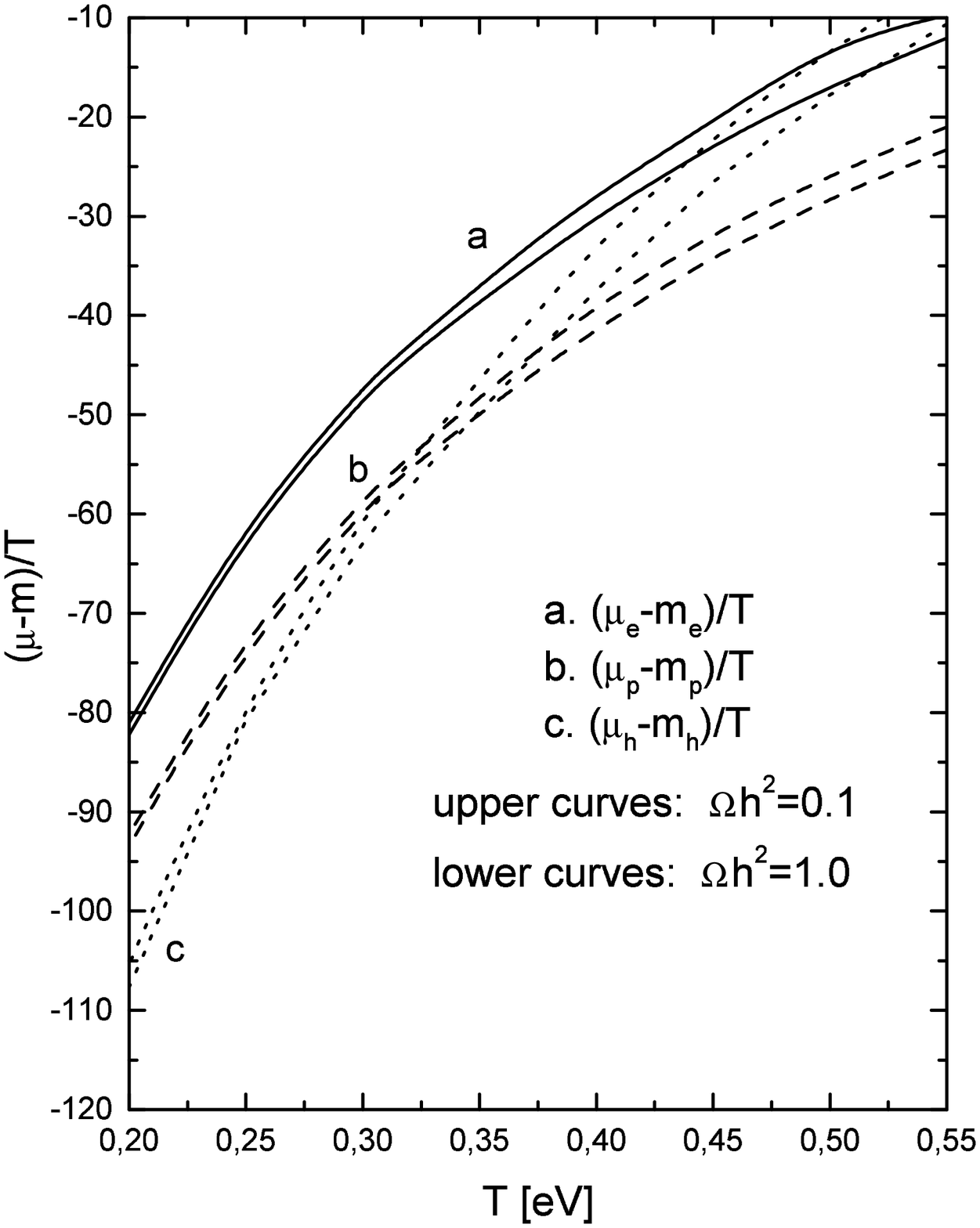} \caption{Left: Standard fraction $X^e_{st}$ used
to compute the corrections in the Saha law. It is worth noticing
how quickly recombination is accomplished. This figure stands for
a standard result, and it is quoted here just for the ease of the
discussion. Right: Evolution of the standard chemical potentials
as a function of the temperature for different values of $\Omega
h^2$. As far as we know, although this figure is an standard
result, it was not presented in cosmology textbooks before. The
numerical results contained in both of these plots were used to
compute non-extensive corrections in the Saha law.}
\end{figure}

Looking at Eq. (\ref{gsl}) we may note that we are going to
confront a problem that does not appear in the standard case: The
chemical potentials appear in a separate way, in such a way that
not only the relationship $\mu_H-\mu_e + \mu_p=0$ enters in the
computation. Since it is not common to find the $\mu$-values, or
even the method to find them, in standard cosmology books, we
shall briefly outline here how they can be naively obtained. We
shall then obtain the standard chemical potentials $\mu=\mu_{st}$,
and consider that generalized expressions for them will be written
as $\mu_q=\mu_{st} + (q-1) C$. Since the term including the factor
$C$ will provide a second order correction in Eq. (\ref{gsl}) we
shall be concerned only with the standard computation. We have to
consider a system of three equations:
\begin{enumerate}
  \item chemical equilibrium $\mu_H-\mu_e + \mu_p=0$
  \item electrical neutrality $n^e=n^p$
  \item the standard result for $1-X^e_{st}/(X^e_{st})^2$.
  The fraction $X^e_{st}$ is shown in Fig.~3.
\end{enumerate}
These three equations form a complicated system but with three
unknowns, regarded here as $(\mu_i - m_i)/T$, where $i$ runs over
the three species involved. Although this system cannot be solved
analytically, it can be solved in a numerical way. The exact
values of the solution will ultimately depend on the value of
$\eta$, the baryon to photon ratio, which in turns depends on the
parameter $\Omega h^2$. Results for the chemical potentials are
shown in Table \ref{che} and in Fig.~3.

\begin{table}[t]
\centering \caption{Standard chemical potentials used in the
determination of the correction at recombination epoch. Two
particular values for $\Omega h^2$ are shown.}
\begin{tabular}{|llllll|}
 \hline T [eV]  & $\frac{\mu_e - m_e}{T}$ & $\frac{\mu_p - m_p}T$ &
 $\frac{\mu_h - m_h}T$ &  $X_e^{st}$ & $\Omega h^2 $\\ \hline
0.55 & -9.7450 & -21.018 & -6.0360 & 1.  & 0.1 \\ 0.50 & -12.652 &
-25.820 & -13.168 &  1. & \\ 0.45 & -20.434 & -31.707 & -21.918 &
0.9998 & \\ 0.40 & -27.808 & -39.082 & -32.890 & 0.9959 & \\ 0.37
& -33.156 & -44.429 & -40.829 & 0.9481 & \\ 0.35 & -37.044 &
-48.317 & -46.504 & 0.7537 & \\ 0.33 & -40.991 & -44.429 & -40.829
& 0.9481 & \\ 0.32 & -43.011 & -54.284 & -54.795 & 0.2305 &\\ 0.31
& -45.107 & -56.380 & -57.616 & 0.1267 & \\ 0.30 & -47.310 &
-58.583 & -60.560 & 0.0647 & \\ 0.25 & -60.806 & -72.077 & -78.486
& 0.0008 & \\ \hline \hline 0.55 & -12.047 & -23.320 & -10.641 &
1. & 1.
\\ 0.50 & -16.850 & -28.123 & -17.773 & 0.9999 &  \\
0.45 & -22.735 & -34.008 & -26.522 & 0.9989 & \\ 0.40 & -30.076 &
-41.349 & -37.426 & 0.9619 & \\ 0.37 & -35.169 & -46.442 & -44.854
& 0.7096 & \\ 0.35 & -38.658 & -49.931 & -49.732 & 0.3786&
\\ 0.33& -42.307 & -53.580 & -54.675 & 0.1431 & \\ 0.32 & -44.251
& -55.525  & -57.277 & 0.0797 & \\ 0.31 & -46.304 & -57.577 &
-60.011 & 0.0419 & \\ 0.30 & -48.484 & -59.757
 & -62.908 & 0.0209 & \\
 0.25 & -61.958 & -73.231 & -80.789 & 0.0002 & \\
 0.20 & -82.191 & -93.464 & -107.655 & 0. & \\ \hline
\end{tabular}
\label{che}
\end{table}

With these values, the correction terms in the Saha law, using
Eq.~(\ref{gsl}), can be immediately computed. The results are
shown in Fig.~4, where two different cases for the baryon content
of the universe are studied. As the corrections in the
non-relativistic expression for the number of particles is
proportional to the factors $((\mu_i - m_i)/T)^2$ , we note that
leading order corrections in the period we are interested in are
$\sim(q-1) 50^2 = (q-1) 2500$. These are large factors, and are
the main reason by which all bounds imposed at the time of
recombination, where matter and radiation cease to be in thermal
contact, i.e. using CMBR data, are much stronger than those
imposed at earlier epochs.

\begin{figure}
\centering \leavevmode \epsfxsize=8.5cm \epsfysize=9.5cm \label{c}
\epsffile{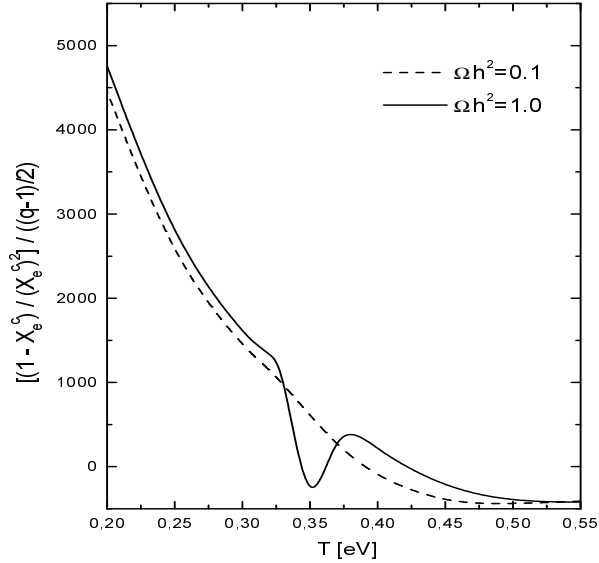} \caption{Final corrections in the Saha law.
Note that the relative strength of the correction differs
appreciably for different baryon densities. Note that the
interesting range for the correction is limited by the
equilibrium in the reaction $D \longleftrightarrow p +n$,
represented by $\mu_d=\mu_n+\mu_p$.}
\end{figure}

\section{The Boltzmann equation}

To analyze in an appropriate way the decoupling of particles we
must consider the microscopic evolution of the distribution
functions $f(p^\mu,x^\mu)$, via the Boltzmann equation. An
interesting recent study on this issue was presented in Ref.
\cite{BARRACO-2}, but their emphasis was put on other aspects. We
shall start here by writing the Boltzmann equation as,
\begin{equation}
\label{eq:LeqC} \hat L[f]=C[f],\end{equation} where $C$ is the
collisional operator and $\hat L$ is the Liouville operator. The
expression for the latter, in the non-relativistic case, for the
phase space density $f(\vec{v},\vec{x})$ of a species with mass
$m$ under a force $\vec{F}=d\vec{p}/dt$, is
\begin{equation}
\hat L=\frac{d}{dt}+\frac{d\vec{x}}{dt} \cdot
\vec{\nabla}_{\vec{x}} + \frac{d\vec{v}}{dt} \cdot
\vec{\nabla}_{\vec{v}},\end{equation} equivalently,
\begin{equation}
\hat L=\frac{d}{dt}+ \vec{v} \cdot \vec{\nabla}_{\vec{x}} +
\frac{\vec{F}}{m} \cdot \vec{\nabla}_{\vec{v}}.\end{equation} The
relativistic generalization becomes,
\begin{equation}
\label{eq:LGR} \hat L=p^\alpha \frac{\partial}{\partial
x^\alpha}-\Gamma^\alpha_{\beta \gamma}p^\beta p^\gamma
\frac{\partial}{\partial p^\alpha}.\end{equation} Gravitational
forces are present through the connection $\Gamma^\alpha_{\beta
\gamma}$. For a Friedmann-Robertson-Walker universe,
$f=f(|\vec{p}|,t)$ (or equivalently $f=f(E,t)$) and the only
non-null components of the connection $ \Gamma^\alpha_{\beta
\gamma}$ are the usual \begin{eqnarray} \Gamma^i_{jk} &=&
\frac{1}{2} h^{il} \left( \frac{\partial h_{lj}}{\partial x^k} +
\frac{\partial h_{lk}}{\partial x^j}  + \frac{\partial
h_{jk}}{\partial x^l}
\right), \\ \Gamma^0_{ij} &=& \frac{\dot R}{R} h_{ij}, \\
\Gamma^i_{0j} &=& \frac{\dot R}{R} \delta^{i}_{\;j},\end{eqnarray}
with $ h^{ij}=-g^{ij}$ and Latin indices running from 1 to 3.

Taken this into account and recalling that $p^\mu=(E,\vec{p})$ and
$x^\mu=(t,\vec{x})$, Eq. (\ref{eq:LGR}) becomes
\begin{equation}
\hat L[f(E,t)]=E \frac{\partial f}{\partial t} - \frac{\dot R}{R}
|\vec{p}|^2 \frac{\partial f}{\partial E}.\end{equation}

Substituting this expression in Eq. (\ref{eq:LeqC}) and taking
into account the definition of the number density, we can
integrate by parts to obtain
\begin{equation}
\dot n + 3 H n=\frac{g}{2\pi} \int C[f]
\frac{d^3p}{E}.\end{equation} We can also rewrite the collisional
term defining
\begin{equation}
\dot f=\frac{g}{2\pi^2} \frac{C[f]}{E},\end{equation} in order to
have the Boltzmann equation as,
\begin{equation}
\label{eq:Boltzdotf} \dot n +3 H n = \int \dot f
d^3p.\end{equation}

The collisional term is usually dominated by annihilations between
particles and anti-particles ($p \bar p$) and we assume now that
the number of particles is identical to the number of
anti-particles. In this way, $\dot f_c$ is given by (see Section
9.2 of Ref. \cite{Peacock} for details):
\begin{equation}
\label{eq:dotf} \dot f_c= - \int <\sigma v> f \bar f d^3\bar
p,\end{equation} where $<\sigma v>$ is the product of the cross
section times the velocity, averaged in velocities space. We can
extract $<\sigma v>$ out of the integral, evaluating it in an
averaged energy.

Let us focus in $\psi$-particles and their anti-particles $\bar
\psi$. Taking into account Eq. (\ref{eq:dotf}) we integrate over
momentum space in Eq. (\ref{eq:Boltzdotf}) to arrive at
\begin{equation}
\dot n_q^\psi + 3H n_q^\psi = -<\sigma v>(n_q^\psi)^2.
\end{equation} Recall that the index $q$ refers to the
non-extensive distribution functions and generalized quantities.
We can now account for the thermal production of particles. Naming
$\Upsilon$ the term giving this contribution, we can write the
Boltzmann equation as
\begin{equation}
\dot n_q^\psi + 3H n_q^\psi = -<\sigma v>(n_q^\psi)^2 + \Upsilon.
\end{equation} This term can be spelled out as in the standard case \cite{Peacock},
invoking thermodynamical
equilibrium: in an static universe $(\dot R =H=0)$, $n_q^\psi$
would be a constant and equal to $n_{q,eq}^\psi$; then,
\begin{equation}
\dot n_q^\psi + 3H n_q^\psi = -<\sigma v> \left( (n_q^\psi)^2
-(n^\psi_{q,eq})^2\right) . \end{equation} We can rewrite the
previous equation in terms of the variable $Y_q\equiv
n_q^\psi/s_q$, where  $Y_{q,eq}\equiv n_{q,eq}^\psi/s_q$,
resulting,
\begin{equation}
\dot n_q^\psi + 3H n_q^\psi = -<\sigma v> s_q^2 \left( Y_q^2
-Y_{q,eq}^2 \right). \end{equation}

All this resembles the usual derivation, and indeed we can go
still further. We can write $\dot n_q^\psi + 3H n_q^\psi$ in terms
of $Y_q$. We shall take into account that $s_q R^3=constant$, and
we derive with respect to time both the definition for $Y_q$ and
the equation $s_q R^3=constant$. Finally recalling that $H=\dot
R/R$, we get
\begin{equation}
\dot n_q^\psi + 3H n_q^\psi=s_q  \dot Y_q. \end{equation}
Combining the last two expression for $\dot n_q^\psi + 3H
n_q^\psi$, we obtain:
\begin{equation}
\label{eq:Yq1} \frac{dY_q}{dt}= -<\sigma v> s_q \left( Y_q^2
-Y_{q,eq}^2 \right). \end{equation}

Introducing explicitly the temperature, via $x=m/T$, with $m$
being any appropriate mass scale, and using the
radiation-dominated-era relation between  $x$ and $T$, Eq.
(\ref{eq:tTrel}), we get
\begin{equation}
\frac{dt}{dx}= 2(0.301)(g_{\star}^q)^{-1/2} \frac{m_{pl}}{m^2} x,
\end{equation}
and we can write
\begin{equation}
\dot Y_q=\frac{dY_q}{dt}=\frac{dx}{dt}
\frac{dY_q}{dx}.\end{equation}

Using the Jacobian $dt/dx$, we can rephrase for $\dot Y_q$
\begin{equation}
\label{eq:Yq2} \dot Y_q=\frac{H(m)}{x}
\frac{dY_q}{dx},\end{equation} where we have defined
\begin{equation}
H(m)\equiv 1.66 (g_{\star}^q)^{1/2} \frac{m^2}{m_{pl}}.
\end{equation}
Note that $H(m)$ is related with $H$ by the simple expression
\begin{equation}
H= x^{-2}H(m).\end{equation}

Finally, using Eqs. (\ref{eq:Yq1}) and (\ref{eq:Yq2}) we can write
the Boltzmann equation as
\begin{equation}
\frac{dY_q}{dx}= -\frac{x <\sigma v> s_q}{H(m)} \left( Y_q^2
-Y_{q,eq}^2 \right), \end{equation} where $<\sigma v>$ is given by
\begin{equation}
<\sigma v>= -\frac{1}{(n_q^\psi)^2} \int \dot f_c
d^3p.\end{equation} Considering other $\psi \bar \psi$
annihilation channels, say $\psi \bar \psi$ into a final state
$F$, an additional term would appear that would be similar to that
in the rhs of the previous equation, but with $<\sigma_{\psi \bar
\psi\rightarrow x \bar x} v>$ replaced by $<\sigma_{\psi \bar
\psi\rightarrow F} v>$. Sum over all annihilation channels yields
the final result in terms of the effective cross section
$<\sigma_A v>$,
\begin{equation}
\frac{dY_q}{dx}= -\frac{x <\sigma_A v> s_q}{H(m)} \left( Y_q^2
-Y_{q,eq}^2 \right). \end{equation} Defining
$\Gamma_A=n_{q,eq}^\psi <\sigma_A v>$, we get
\begin{equation}
\label{eq:Beqf} \frac{x}{Y_{q,eq}} \frac{dY_q}{dx} = -
\frac{\Gamma_A}{H(x)} \left[ \left(\frac{Y}{Y_{eq}}\right)^2 -1
\right]. \end{equation}\\

{\it Remark 8: Note that this derivation is independent of the
particular form for the distribution functions, but the final
product has the same aspect than that obtained in the standard
scheme, see for instance Section $(5.2)$ of Ref. \cite{Kolb}.}

\subsection{Freezing}

Following the standard procedure, we would like to study how much
the freezing times and temperatures are modified by a change in
the statistical description. Using Eq. (\ref{eq:Beqf}), we see
that the comoving number density of $\psi$ particles is controlled
by the ratio $\Gamma/H$, and the amount of the deviation of the
distribution functions from their equilibrium values. When
$\Gamma/H$ is less than unity, the change in the number density is
small, $-\Delta Y_q/Y_q\sim -(xdY/dx)/Y_{q,eq}\sim \Gamma/H < 1$,
annihilations stop and the number density freezes.

The annihilation rate $\Gamma_A$ varies as $n_{q,eq}$ times the
averaged cross section  $<\sigma_A v>$. In the relativistic
regime, $n_{q,eq} \propto T^3$. In the non-relativistic regime,
$n_{q,eq} \propto e^{-m/T}$, and $\Gamma_A$ exponentially
decreases. In both regimes, $\Gamma_A$ decreases with temperature,
and eventually goes below the necessary rate to maintain
equilibrium, which by definition occurs when  $x=x_f$ (the
``freeze out''). Then we expect that for $x<x_f$, $Y_q(x)\simeq
Y_{q,eq}(x)$, whereas for $x>x_f$, $Y_q(x>x_f)=Y_{q,eq}(x_f)$.

We can compute the value of $Y_{q,eq}\equiv n_{q,eq}^\psi/s_q$ in
each of the previous cases recalling the definition  of $s_q$ and
Eqs. (\ref{eq:nqb}) and (\ref{eq:nqf}) for relativistic ($x\ll 3$)
particles and Eq. (\ref{nr}) for non-relativistic ($x\gg 3$)
particles. We obtain:
\begin{equation}
Y_{q,eq}^b = 0.278 g^b \left[ 1 + 6
\frac{\zeta(4)}{\zeta(3)}(q-1)\right] (g_{\star
s}^q)^{-1}\end{equation} for relativistic bosons,
\begin{equation}
Y_{q,eq}^f = 0.278 \frac{3}{4} g^f (g_{\star s}^q)^{-1}\left[ 1 +
8 \frac{\zeta(4)}{\zeta(3)}(q-1)\right] \end{equation} for
relativistic fermions, and
\begin{equation}
Y_{q,eq}^{nr} = 0.145 g x^{3/2} e^{-x} (g_{\star s}^q)^{-1} \left[
1+ \frac{q-1}{2} \left( \frac{15}{4}+3x+x^2 \right)
\right]\end{equation} with $x\gg 3$, for non-relativistic
particles.

\begin{figure}[t]
\begin{center}
\vspace{-2.5cm}
\includegraphics[width=10cm,height=13cm]{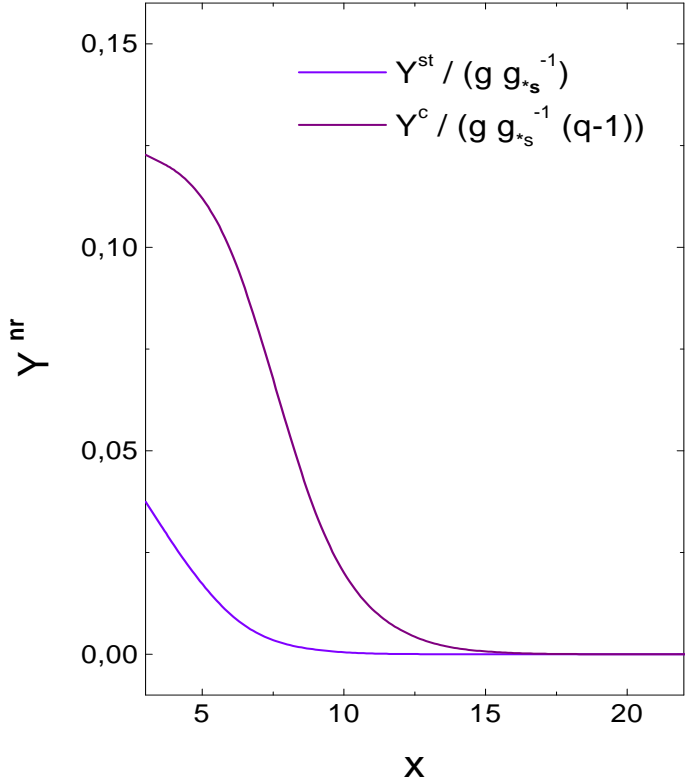}
\end{center}
\vspace{-1.5cm} \caption{Standard equilibrium abundance and
non-extensive correction for non-relativistic particles. }
\label{fig:Ynr}
\end{figure}

Since we can write $g^q_{\star s}$ as $g^q_{\star s}=g^{st}_{\star
s}+ (q-1)\hat g^c_{\star s}$, we can get a first order result for
the above mentioned quantities,
\begin{equation}
\label{eq:Yqeqrb} Y_{q,eq}^b = 0.278 g^b (g_{\star s}^{st})^{-1}
\left[ 1 + (q-1)\left(6 \frac{\zeta(4)}{\zeta(3)}-\frac{\hat
g^c_{\star s}}{g^{st}_{\star s}}\right)\right],\end{equation}
\begin{equation}
\label{eq:Yqeqrf} Y_{q,eq}^f = 0.278 \frac{3}{4} g^f (g_{\star
s}^{st})^{-1} \left[ 1 +(q-1)\left(8
\frac{\zeta(4)}{\zeta(3)}-\frac{\hat g^c_{\star s}}{g^{st}_{\star
s}}\right)\right],\end{equation} and
\begin{equation}
\label{eq:Yqeqnr} Y_{q,eq}^{nr} = 0.145 g x^{3/2} e^{-x} (g_{\star
s}^{st})^{-1}\left[ 1+ \frac{q-1}{2} \left( \frac{15}{4}+3x+x^2
-2\frac{\hat g^c_{\star s}}{g^{st}_{\star s}}\right) \right]
\end{equation} with $x\gg 3$.

In Fig.~5, we show the equilibrium abundance for non-relativistic
particles together with the non-extensive correction. The
correction diminishes exponentially with decreasing temperature,
but this is an effect of the overall factor $e^{-x}$ also present
in the standard result, and not of the non-extensivity introduced.

\section{Current values}

We are interested in the value of the corrected ratio
\begin{equation}
\Omega_q= \frac{\rho_q }{\rho_c} \qquad \textrm{with} \quad
\rho_c=\frac{3H_0^2}{8\pi G},\end{equation} where $G=m_{pl}^{-2}$
with $m_{pl}=1.2211\times 10^{19}$ GeV and $H(t)$ is $H_0=2.1332 h
\times 10^{-42}$ GeV $(0.4<h<1)$. From the current value of
$\rho_q^R$, we can get the contribution of relativistic particles
\begin{equation}
\Omega_{q,R}h^2=\frac{8\pi G}{3H_0^2} h^2\rho_q^R.\end{equation}
To get $\rho^R_q$ we must know the the value of $g_{\star}^q$ and
$g_{\star s}^q$.

Carrying out a first order computation using Eqs. (\ref{eq:gsst}),
and (\ref{eq:gsc}), and Eqs. (\ref{a}) and (\ref{b}), and assuming
a universe populated by photons and three types of neutrinos, we
get
\begin{equation}
\label{eq:gsqh} (g_\star^q)_{today}= 3.36 \left[1 + 9.43
(q-1)\right]\end{equation}
\begin{equation}
\label{eq:gssqh} (g_{\star s}^q)_{today} =3.91 \left[1 +
7.11(q-1)\right]\end{equation}

Taking the current value of the photon temperature, $T=2.728 \pm
0.004 $ K, we can obtain: $\rho_q^R$, $s_q$, $n_q^\gamma$ and
$\Omega_{q,R}$,
\begin{eqnarray}
(\rho_q^R)_{today} &=& 7.83 \times 10^{-34}\left[ 1 + 9.43
(q-1)\right] {\rm g\;cm^{-3}},\\ (s_q)_{today} &=& 2899.41 \left[
1 + 7.11 (q-1)\right]{\rm cm^{-3}},\\ (n_q^\gamma)_{today} &=&
411.77
\left[ 1 + 5.4 (q-1)\right]{\rm cm^{-3}},\\
\label{eq:OmqR}(\Omega_{q,R})_{today} &=& 4.17 \times
10^{-5}\left[ 1 + 9.43 (q-1)\right]h^{-2}.
\end{eqnarray}

\section{Wimps in non-extensive statistics}

We now consider weakly interacting massive particles, for instance
neutrinos with very small mass, in non-extensive statistics.

\subsection{Relativistic relics, $x_f <  3$}

In this case, $Y_{q,eq}$ is not changing with time, (see Eqs.
(\ref{eq:Yqeqrb}) and (\ref{eq:Yqeqrf})). The asymptotic value of
$Y_q$, ($Y_q(x\rightarrow \infty)$) $\equiv Y_{q,\infty}$, is just
the equilibrium value at freezing:
\begin{equation}
Y_{q,\infty}=Y_{q,eq}^f(x_f) = 0.278 \frac{3}{4} g^f \left[ 1 + 8
\frac{\zeta(4)}{\zeta(3)}(q-1)\right] (g_{\star s}^q(x_f))^{-1}
\qquad \mbox{\textrm{with} } \qquad x_f < 3.
\end{equation}
The current abundance of the  $\psi $ species today is
\begin{equation}
n_{q, today}^\psi=s_{q, today} Y_{q, \infty}, \end{equation}
\begin{equation}
n_{q, today}^\psi=\frac{3}{4} g^f 806.04 \left[ 1 +
14.31(q-1)\right] (g_{\star s}^q)^{-1}. \end{equation} The mass
density contribution of this kind of hot relics is given by
\begin{equation}
\rho_{q, today}^\psi = n_{q, today}^\psi m_\psi, \end{equation}
and the fraction of the critical mass will be
\begin{equation}
\Omega_{q, today}^\psi h^2= \frac{8\pi G}{3H_0} h^2 \rho_{q,
today}^\psi = \Omega_{q, today}^\psi h^2= \frac{3}{4} g^f 7.65
\times 10^{-2} \left[1 + 14.31(q-1) \right] (g_{\star s}^q)^{-1}
\left( \frac{m_\psi}{{\rm eV}} \right).\end{equation}

We know that $\Omega_0 h^2 < 1$; then we can apply this bound to
the contribution to the species $\psi$ to $\Omega_0 h^2$ and
obtain a cosmological upper limit to the mass of this kind of
particles
\begin{equation}
m_\psi < 13.08 \frac{4}{3}\left[1 - 14.31(q-1) \right] g_{\star
s}^q (g^f)^{-1}.\end{equation} Light neutrinos would decouple when
$T \sim$ MeV. At this temperature
\begin{equation}
g_{\star s}^q(x_f)= 2 + \frac{7}{8} \left[2+2+ 2\times 3\right] +
\frac{60}{2\pi^2}\frac{45}{\pi^2} \frac{\zeta(5)}{2} (q-1)\left[2+
\frac{15}{16} (2+ 2\times 3)\right],\end{equation}
\begin{equation}
g_{\star s}^q(x_f)=10.75 \left[1+ 7.60(q-1)\right].\end{equation}
Then, for an species with two components, $g^f=g^\nu=2$, and to
first order in $(q-1)$, it results
\begin{equation}
\label{eq:mmubound } m_\nu < 93.72 \left[1 - 6.7(q-1) \right] {\rm
eV}.\end{equation} This is a non-extensive generalization of the
Cowsik and McClelland bound, see for instance Ref. \cite{Kolb}.
Even when is improbable that an experiment with enough sensitivity
could differentiate among values of $q$ using Eq.
(\ref{eq:mmubound }), further recalling that it is an upper bound
what that equation is imposing, it is worth noticing that a change
of statistics can have a direct influence upon the mass spectrum
of the particles.

\subsection{Cold relics, $x_f> 3$}

In this case,
\begin{equation}
\label{eq:Yqcold} Y_{q,eq}^{nr}(x_f) = 0.145 g x_f^{3/2} e^{-x_f}
\left[ 1+ \frac{q-1}{2} \left( \frac{15}{4} + 3x_f + x_f^2 \right)
\right] (g_{\star s}^q)^{-1}.\end{equation} Now,  $Y_{q,eq}$
strongly depends on $m$ (since $x_f=m/T_d$). To make a numerical
estimation we have to determine $T_d$, using the condition $\Gamma
\simeq H$. Reactions able to change the number of wimps  $A$, are
like $A \bar A\leftrightarrow X \bar X$, where $X$ is a generic
particle, assumed in thermal equilibrium. The average value of
$\sigma v$ can be expressed as
\begin{equation}
<\sigma v>\equiv \sigma_0 \left(\frac{T}{m}\right)^k.
\end{equation} The value of the power law index $k$ will depend on
the details of the annihilation process, usually it is of order 1.
The value of $\sigma_0$ depends on $m$ and has a simple form for
the extreme cases $m\ll m_Z$ and $m\gg m_Z$, where $m_Z\simeq
10^2$ Gev is the mass of the $Z$-boson.

For wimps with $m<m_Z$, the effective cross section $\sigma_0$ can
be written as \cite{Padma},
\begin{equation}
\label{eq:sigma0} \sigma_0\simeq \frac{c}{2\pi}G_F^2
m^2,\end{equation} where $G_F$ is the Fermi constant,
$G_F^{-2}=292.8$ GeV. The value of $c$ depends on the fermion. We
shall consider Dirac (spin 1/2) particles for which  $c\simeq 5$.

The reaction rate is given by
\begin{equation}
\Gamma= n_q<\sigma v>, \end{equation} or, equivalently, using Eq.
(\ref{nr}),
\begin{equation}
\Gamma = \frac{\sigma_0 g_A}{(2\pi)^{3/2}} T^3 \left(\frac{m}{T}
\right)^{3/2-k} e^{-m/T} \left[ 1 + \frac{q-1}{2} \left( \frac
{15}4 + 3 \frac {m}{T} + \left(\frac {m}{T}\right)^2 \right)
\right].\end{equation} The expansion rate $H$ as a function of
temperature is
\begin{equation}
H =1.66 \left( g_*^q \right)^{1/2} \frac{T^2}{m_{Pl}}.
\end{equation} Then, the condition $\Gamma/H\simeq 1$ yields
\begin{equation}
\label{eq:Gamma/H=1} 1=3.82\times 10^{-2} g_A (g_{\star}^q)^{-1/2}
\left(\frac{m}{T_d} \right)^{1/2-k} e^{-m/T_d} \left[ 1 +
\frac{q-1}{2} \left( \frac {15}4 + 3 \frac {m}{T_d} + \left(\frac
{m}{T_d}\right)^2 \right) \right] \sigma_0 m m_{pl}.\end{equation}
Solving for $e^{-m/T_d}$, substituting the result in Eq.
(\ref{eq:Yqcold}) for $Y_{q,eq}^{nr}$, and computing $\sigma_0$
using Eq. (\ref{eq:sigma0}), we get
\begin{equation}
\label{eq:Yqeqcr} Y_{q,eq}^{nr}(x_f)= 2.86\times 10^{-9}
(g_{\star}^q)^{-1/2}
\left(\frac{m}{T_d}\right)^{k+1}\left(\frac{m}{{\rm
GeV}}\right)^{-3}.\end{equation} Given the current value
$s_{q,today} = 2899.41$  $\times \left[1 + 7.11 (q-1) \right]
cm^{-3}$, we obtain
\begin{equation} n_q=8.3\times 10^{-6} (g_\star^q)^{-1/2}
\left(\frac{m}{T_d}\right)^{k+1}\left(\frac{m}{{\rm
GeV}}\right)^{-3} \left[ 1 + (q-1) \left[ 7.11 + \frac{1}{2}
\left( \frac {15}4 + 3 \frac {m}{T_d} + \left(\frac
{m}{T_d}\right)^2 \right) \right] \right] .\end{equation} The
energy of these particles today is $\rho_{q,today}=n_{q,today}m$,
and the contribution to the critical density is \ben
\Omega_{q,today}h^2= \frac{8\pi G}{3H_0}h^2 \rho_{q,today}=
\hspace{10cm} \nonumber \\  0.79 (g_\star^q)^{-1/2}
\left(\frac{m}{T_d}\right)^{k+1}\left(\frac{m}{{\rm
GeV}}\right)^{-3} \left[ 1 + (q-1) \left[ 7.11 + \frac{1}{2}
\left( \frac {15}4 + 3 \frac {m}{T_d} + \left(\frac
{m}{T_d}\right)^2 \right) \right] \right].\een

We would now like to solve for $T_d$. Taking logarithm in Eq.
(\ref{eq:Gamma/H=1}), we obtain \ben \label{eq:m/Td}
\frac{m}{T_d}=  17.74 + \hspace{12cm}\nonumber \\
\ln\left[\frac{g_A}{(g_\star^q)^{1/2}}\right] + \left(\frac{1}{2}
- k\right) \ln\left[\frac{m}{T_d}\right] + 3
\ln\left[\frac{m}{{\rm GeV}}\right] + \ln \left[ 1 + (q-1) \left[
\frac{1}{2} \left( \frac {15}4 + 3 \frac {m}{T_d} + \left(\frac
{m}{T_d}\right)^2 \right) \right] \right] .\een Since $g_\star^q$
is a slowly varying function we can solve this equation in an
iterative form. Consider the case of a wimp with $m> 1$ GeV. From
Eq. (\ref{eq:m/Td}) we see that $m/T_d \simeq 17.74$, assuming for
simplicity that $k=0$, the term $\ln(m/T_d)$ corrects the value
$m/T_d$ to  $m/T_d\simeq 19.18$, giving $T_d \simeq 52$ MeV
($m$/GeV). At this temperature, all species are relativistic and
all of them contribute to $g_\star^q$, making it of the order of
100. Taking this into account, the term
$\ln((g_\star^q)^{1/2}/g_A)\simeq \ln5 \simeq 1.61$, and the value
of $m/T_d$ gets corrected to $m/T_d\simeq 17.57$. The
non-extensive term  is
\begin{equation}
\ln \left[ 1 + (q-1) \left[ \frac{1}{2} \left( \frac {15}4 + 3
\frac {m}{T_d} + \left(\frac {m}{T_d}\right)^2 \right) \right]
\right] .\end{equation} Substituting this expression in the zero
order of  $m/T_d$ we get \begin{equation} \frac{m}{T_d} = 17.57 +
3\ln \left(\frac{m}{{\rm GeV}}\right) + (q-1) 185.8.\end{equation}
For masses $m \sim 1$ GeV,
\begin{equation}
\frac{m}{T_d} = 17.57 + (q-1) 185.8.
\end{equation}

The use of this result in Eqs. (\ref{eq:Yqeqcr}) yields
\begin{equation}
Y_{q,eq}^{nr}=4.87 \times 10^{-9} \left[1 + 10.5 (q-1)
\right]\left( \frac{m}{{\rm GeV}}\right)^{-3},\end{equation}
\begin{equation}
\Omega_{q,today}h^2= 1.34 \left[1 - 75.2 (q-1) \right]\left(
\frac{m}{{\rm GeV}}\right)^{-2}.\end{equation} The fermion $A$ and
its anti-particle $\bar A$, will provide twice this value of
$\Omega_{q,today}h^2$, i.e. the constraint $\Omega_{q,today}h^2 <
1$ is then a corrected lower bound for the mass of these particles
\begin{equation}
m > 1.64 \left[1 - 37.6 (q-1) \right] \rm{GeV}.\end{equation}

\section{Modification of the matter-radiation equality}

Let us characterize the matter-radiation equality by the time of
its occurrence $t=t_{eq}$, its scale-factor $R=R_{eq}$, and its
redshift $z=z_{eq}$. From the equalities
\begin{eqnarray}
\rho_{matter}(t_{eq})&=&\rho_{matter}(t_0) \left(
\frac{R(t_0)}{R(t_{eq})}\right)^3=\rho_c
\Omega_{matter}(t_0)(1+z_{eq})^3, \\
\rho_{radiation}(t_{eq})&=&\rho_{radiation}(t_0) \left(
\frac{R(t_0)}{R(t_{eq})}\right)^4=\rho_c
\Omega_{matter}(t_0)(1+z_{eq})^4, \end{eqnarray} it follows
\begin{equation}
1+z_{eq}=\frac{\Omega_{matter}(t_0)}{\Omega_{radiation}(t_0)}\simeq
\frac{\Omega}{\Omega_{radiation}(t_0)}. \end{equation} And then,
\begin{equation}
1+z_{eq}= (\Omega_R(t_0) h^2)^{-1} \Omega h^2.\end{equation} In
the non-extensive setting,
\begin{equation}
1+z_{eq}= 2.4\times 10^4\left [ 1 - 9.43 (q-1)\right] \Omega
h^2\end{equation} corresponds to a temperature given by
$T_{eq}=T_0 (1+z_{eq})$. Taking $T_0=2.728$~K, we obtain
\begin{equation}
T_{eq}=5.64\left [ 1 - 9.43 (q-1)\right] \Omega h^2,\end{equation}
and for the time we get
\begin{equation}
t_{eq}\simeq 0.39 H_0^{-1} \Omega^{-1/2} (1+z_{eq})^{-3/2}=
1.022\times 10^3\left [ 1 + 14.14 (q-1)\right] (\Omega h^2)^{-2}
\rm{years}.\end{equation}

\section{Conclusions}

In this paper we have studied how the statistical description that
we suppose is valid in the early moments of the universe affects
different processes along the thermal history of our basic
cosmological model. In particular, starting from the modifications
on the energy and number densities, we explored all consequences
introduced by non-extensivity in the processes of decoupling,
particle--anti-particle excess, recombination, photon-neutrino
temperature relationship, matter-radiation equality, and others.

In order to avoid re-stating our obtained results, let us just
mention a general conclusion. Suppose that all cosmological
observables are measured with a precision of 1\%. Suppose, too,
that the standard values for these observables agree well with
observations, i.e. the standard values are always within the error
bar of the experiments. (This latter assumption is not always
true, particularly when more than one experiment is involved at
the same time, like in the case of the determination of the baryon
density using nucleosynthesis and recent CMBR measurements.) When
relativistic particles are involved, we can summarize our results
as follows. Given any standard prediction, the correction
introduced by non-extensivity is of the order of ten times the
standard value,
\begin{equation}
X=X_{st} + (q-1) \times F,
\end{equation}
where the factor $F$ is order 10$X_{st}$. Then, given the previous
assumptions, when relativistic particles are involved, $|q-1|$ is
bound to be of the order $10^{-3}$. On the other hand, when
non-relativistic particles are involved, the corrections are much
bigger, and $F$ can be at least of the order of 1000$X_{st}$. In
these cases, and also within the previous assumptions, $|q-1|$ is
bound to be less than $10^{-4}$. It is worth noticing that this
simple conclusion is sustained by all currently existing bounds,
and it is also worth noticing that the ultimate reasons by which
the constraints on $|q-1|$ are more restrictive in the epoch of
recombination than those obtained in nucleosynthesis (see Paper
II) can be tracked from this formal point of view.

In our second paper in this series we shall study the
nucleosynthesis epoch, and will provide an analytical assessment
of the Helium 4 and Deuterium primordial production in the
framework of non-extensive statistics, including corrections
coming from the free neutron decay process. We shall particularly
focus on how non-extensivity affects the principle of detailed
balance.

\subsection*{Appendix: Useful mathematical formulae}


It is convenient to introduce the following functions:
\begin{equation}
e^x_q\equiv [1+(1-q)x]^{1/(1-q)}, \qquad \forall(x,q)
\end{equation}
supplemented by the definition, for  $q<1$, $e^x_q=0$ if
$1+(1-q)x\leq 0$, and for $q>1$, $e^x_q$ diverges in $x=1/(q-1)$,
and
\begin{equation}
\ln_q x \equiv \frac{x^{1-q}-1}{1-q}, \qquad \forall(x,q).
\end{equation}
It is immediate to show that $\lim_{q\rightarrow 1} e^x_q=e^x$ and
that  $\lim_{q\rightarrow 1} \ln_q x=\ln x$. Also the following
properties are valid
\begin{equation}
e_q^{\ln_q x}=\ln_q e_q^x=x, \qquad \forall(x,q).
\end{equation}
With these definitions,
\begin{equation}
S_q=k\ln_q W.
\end{equation}

\subsection*{Acknowledgments}

M. E. Pessah is supported by Fundaci\'on Antorchas. He thanks A.
Platzek for advice. D.F.T thanks A. Lavagno, G. Lambiase, and P.
Quarati for valuable discussions. He was supported by CONICET as
well as by funds granted by Fundaci\'on Antorchas, and
acknowledges the hospitality provided by the Politecnico di
Torino, the University of Salerno, and the ICTP (Italy) during
different stages of this research.


\end{document}